\documentclass[a4paper,11pt]{article}
\usepackage{jcappub} 
\usepackage[T1]{fontenc} 
\usepackage{diagbox}
\usepackage{multirow}
\usepackage{amsmath}
\usepackage{amssymb}
\usepackage{smartdiagram}
\usesmartdiagramlibrary{additions}
\usepackage{booktabs}
\usepackage{tabularx}
\usepackage{soul}
\usepackage{subcaption}
\usepackage{graphicx}
\usepackage{lipsum}
\usepackage{comment}
\usepackage{xcolor}
\usepackage{hyperref}
\usepackage[normalem]{ulem}
\usepackage{bigints}
\usepackage[ruled]{algorithm2e}
\usepackage{multicol}
\usepackage{cancel}
\usepackage{siunitx}
\usepackage{mathtools}
\usepackage{cuted}
\usepackage[toc,page]{appendix}
\usepackage{placeins}
\usepackage{float}
\usepackage{capt-of}
\usepackage{adjustbox}

\pdfoutput=1

\usepackage{aas_macros}

\newcommand{\MG}[1]{\textcolor{black}{#1}}
\newcommand{\GM}[1]{\textcolor{black}{#1}}

\def\permill{\ensuremath{{}^\text{o}\mkern-5mu/\mkern-3mu_\text{oo}}}

\title{\boldmath{\fontsize{22}{11}\selectfont \textcolor{black}{Modelling the next-to-leading order matter three-point correlation function using FFTLog}}}

\author[a, b]{M. Guidi}
\author[a, b]{A. Veropalumbo,}
\author[a, b, c]{E. Branchini,}
\author[d]{A. Eggemeier,}
\author[e]{C. Carbone}

\affiliation[a]{Dipartimento di Fisica, Universit\`a di Roma Tre, Via della Vasca Navale 84, I-00146 Roma, Italy}
\affiliation[b]{INFN - Sezione di Roma Tre, Via della Vasca Navale 84, I-00146 Roma, Italy}
\affiliation[c]{Dipartimento di Fisica, Universit\`a degli Studi di Genova, and INFN Sezione di Genova, Via Dodecaneso 33, I-16146, Genova, Italy}
\affiliation[d]{
Argelander Fellow; Argelander Institut für Astronomie der Universität Bonn, Auf dem Hügel 71, 53121 Bonn, Germany}
\affiliation[e]{INAF -- Istituto di Astrofisica Spaziale e Fisica cosmica di Milano (IASF-MI), Via Alfonso Corti 12, I-20133 Milano, Italy}

\emailAdd{massimo.guidi@uniroma3.it}

\abstract{ 
The study of higher-order statistics, particularly 3-point statistics, of the Large Scale Structure (LSS) of the Universe 
provides us with unique information on the biasing relation between luminous and dark matter and on deviations from primordial Gaussianity. As a result, much effort has been put into improving 
measurement techniques as well as theoretical modelling, 
especially in Fourier space. Comparatively, little progress has been made, instead, in configuration space analyses.  This work represents a first step towards filling this gap by proposing a new strategy for modelling 3-point statistics at higher perturbative orders in configuration space. Starting from the next-to-leading order model for the matter bispectrum, we use 2D-FFTLog to generate its counterpart in configuration space. We calibrate the procedure using the leading order predictions for which an analytic model for the 3-point correlation function (3PCF) already exists. Then we assess the goodness of the 3PCF model by comparing its predictions with measurements performed on the matter distribution in collisionless cosmological N-body \GM{simulation (DEMNUni)}. We focus on two redshifts ($z=0.49$ and $z=1.05$)  in the range spanned by  current and future galaxy redshift surveys. The $\chi^2$ analysis reveals that the next-to-leading order 3PCF models significantly improve over the leading order one for all triangle configurations in both redshifts, increasing the number of matched configurations at redshift $z= 1.05$ and $z= 0.49$, respectively. In particular, a significant improvement is also seen on the Baryonic Acoustic Oscillations (BAO) scale for triangle configurations whose smallest side length is well into the nonlinear regime. The computational cost of the model proposed here is high but not prohibitively large --- order of five hours in a 48-core computation --- and represents the first step towards a complete 3PC model for the galaxies.

 } 

\keywords{
cosmology: large-scale structure of Universe – theory – methods: statistical
}

\begin{document}
\label{firstpage}

\sloppy

\maketitle

\newpage
\newpage

\section{Introduction}
\label{sec:intro}
Galaxy Clustering is a cornerstone in supporting and investigating the validity of the standard cosmological model. So far, two-point statistics have emerged and will remain the primary probes in constraining the physical properties of the universe as long as cosmological fields obey Gaussian statistics.
If this is not the case, additional, fundamental information is captured by higher-order statistics.
The three-point correlation function (3PCF) and its Fourier counterpart, the bispectrum, are the tools with the highest signal-to-noise to characterise departures from Gaussianity and, in so doing, to investigate primordial non-Gaussianity (PNG) and galaxy biasing. Furthermore, combined with two-point statistics, they can break degeneracy among fundamental cosmological parameters. For these reasons, three-point statistics are now a long tradition in studying the Large Scale Structure of the universe (LSS) \citep{Peebles, Scoccimarro2001, Takada_2003, Gaztanaga_2005, Pan, Marin_2008, McBride2011, SlepianEisenstein2017}. 

They play and will play a fundamental role in the exploitation of  next-generation datasets generated by upcoming spectroscopic galaxy surveys like Euclid \citep{EuclidReport, Euclid2013}, Dark Energy Spectroscopic Instrument \citep{DESI2016}, Large Synoptic Survey Telescope (LSST) \citep{LSST2018} and Nancy Grace Roman Grace Space Telescope high latitude survey
\citep{Eifler_2021}. 

To efficiently extract cosmological information through three-point statistics from present and future datasets, it is mandatory to obtain reliable theoretical predictions on the largest possible number of triangle configurations, which implies including the numerous triangles of small size, i.e. to 
probe the nonlinear regime \citep{Sefusatti2006}. 

Standard Perturbation Theory (SPT, see \citep{Bernardeau2002} for a review) of clustering statistic
has proved, so far, an effective way of accessing nonlinear scales in Fourier space. 
Hence the widespread use of perturbative expansion techniques to investigate the clustering properties of the matter in the universe through the power spectrum and bispectrum statistics \citep{Scoccimarro97, Scoccimarro98, Sefusatti_2010, Lazanu_2018, Bose_2018, Alkhanishvili:2021pvy}.
Concerning bispectrum, efforts have been made to develop alternative routes by re-summing perturbative contributions in the Eulerian \citep{Crocce2006, Crocce2008, Bernardeau2012}, Lagrangian \citep{Matsubara2008} and Effective Field Theory (EFT) approaches \citep{Carrasco2014, Hertzberg2015, Angulo:2014tfa, Baldauf2015}. The state of 3-point correlation modelling in configuration space is comparatively less advanced.
\MG{So far, 3PCF models in real and redshift space have been developed at the tree level only \citep{SlepianEisenstein2017, Jing2003, Gaztanaga_2005, Kuruvilla2020}}. They have been successfully used to analyse clustering on quasi-linear scales \citep{Moresco2017, Moresco2021, Slepian:2016kfz, Sugiyama:2020uil, Veropalumbo2021}, hence limited to a relatively small fraction of available triplets and
missing information from the mildly nonlinear scales.
This is because modelling 3-point statistics in configuration space is complicated due to the relation with the Fourier space counterpart in which models are provided. The inverse-Fourier transform induces a scale mixing that requires adopting a computationally demanding numerical approach, even in the mildly nonlinear regime. As yet, direct modelling in configuration space has not yet been explored due to the complexity of fluid equations in configuration space.
On the other hand, the 3PCF approach offers a significant advantage when dealing with real datasets consisting of galaxy surveys with complicated geometry. In Fourier space, the survey footprint induces mode coupling in Fourier that requires computationally demanding numerical approaches \cite{Pardede:2022udo, Philcox_2021}. For 3PCF, the impact of the survey footprint can be efficiently corrected at the estimator level.
A second element that has hampered the development of the 3PCF tool has been, until not long ago, the computational cost of the standard estimators counting all triplets in the sample. 
The situation has changed dramatically since new types of 3PCF estimators capable of reducing computational cost from $N^3$ to $N^2$, have been proposed \citep{Slepian_2015, Slepian_2018, Sugiyama_2018}. 
This work aims to fill the theoretical gap with Fourier space-based 3-point correlation models.
For this, we have developed an accurate and efficient way to map second-order perturbative expansion models of the matter bispectrum into 3PCF predictions using the 2D-FFTlog \citep{Fang2020} technique.

The manuscript is organised as follows. In Sec. \ref{sec:overview}, we briefly review the status of the 3PCF model, introducing the Legendre expansion.
In Sec. \ref{sec:models}, we review the status of 3-point modelling in Fourier space, presenting the strategy for modelling 3PCF  from Fourier space predictions, focusing on the 2D-FFTLog algorithm 
we extensively use to generate 3PCF models. In Sec. \ref{sec:measure}, we present the set of simulated datasets and estimators we use to obtain 3PCF measurement that we then compare with 3PCF models
in Sec. \ref{sec:matching}, where we analyse residuals and perform a $\chi^2$ analysis to validate the models.
In Sec. \ref{sec:results}, we present the results of our analysis, and in Sec. \ref{sec:summary} we summarise and discuss the outcome of our analyses and future developments.

\section{Modelling the matter 3PCF}
\label{sec:models}

\subsection{3PCF definition}
\label{sec:overview}
\label{subsec:3pcfdef}
The matter three-point correlation function in real space, i.e. not taking account of redshift space distortions, is defined as

\begin{eqnarray}
    \
    \label{eq:cl_3pcf_def}
     \zeta_m(r_{12}, r_{13}, r_{23}) & = &\left< \delta_m(\mathbf{x}) \delta_m(\mathbf{x}+\mathbf{r}_{12}) \delta_m(\mathbf{x}+\mathbf{r}_{13}) \right>, \, 
     \
\end{eqnarray}
where $\delta_m(\mathbf{x})$ is the matter density
contrast at the position $\mathbf{x}$,
 $\langle...\rangle$ indicates ensemble average, $\mathbf{r}_{12} = \mathbf{r}_{1} -\mathbf{r}_{2}$, $\MG{\mathbf{r}_{13} = \mathbf{r}_{1} - \mathbf{r}_{3}}$ given the generic vectors $\mathbf{r}_{1}$ and $\mathbf{r}_{2}$, and we use the general notation $r_{i} =|\mathbf{r}_i|$. The definition given in Eq. \ref{eq:cl_3pcf_def} depends only on $r_{ij}$ due to the assumption of isotropy in real space.
 
 This definition can be expanded in Legendre polynomials as

\begin{eqnarray}
   \label{eq:leg_resum}
   \zeta_m(r_{12}, r_{13}, r_{23}) = \sum_\ell \zeta_\ell(r_{12}, r_{13}) \mathcal{L}_\ell (\hat{\mathbf{{r}}}_{12} \cdot \hat{\mathbf{{r}}}_{13}),
\end{eqnarray}
where the Legendre transformation gives the coefficients of the expansion
\begin{eqnarray}
   \label{eq:leg_transf}
   \zeta_\ell(r_{12}, r_{13}) = \frac{2l + 1}{2} \int_{-1}^{+1} \mathrm{d \mu} \ \zeta_m(r_{12}, r_{13}, r_{23}) \ \mathcal{L}_\ell(\mu),
\end{eqnarray}
where $\mathcal{L}_\ell(\mu)$ are the Legendre polynomials and $\mu = \mathbf{r_{12} \cdot r_{13}}/{r_{12}r_{13}}$.

\subsection{The strategy}
Modelling the 3PCF at higher perturbative orders represents an important target for extracting cosmological information from future datasets. Still, it suffers from the fact that it has been limited to tree-level computations so far, and next-to-leading order models have not been computed in Lagrangian Perturbation Theory (LPT,  \cite{Matsubara2008}). 

The strategy we adopt here to model 3-point statistics in configuration space exploits the existing theory in Fourier space in an SPT framework and uses inverse Fourier transform to bring it to configuration space:

\begin{equation}
    \begin{split}
    \label{eq:bk_transform}
    \zeta_{m} (r_{12}, r_{13}, r_{{23}})  = (2\pi)^3 \int  \frac{\mathrm{d}^3k_{12} \mathrm{d}^3k_{13} \mathrm{d}^3k_{23}}{(2\pi)^9}\, & B_m(k_{12}, k_{13}, k_{23})\,e^{i\,(\mathbf{k_{12}} \cdot \mathbf{r_{12}} + \mathbf{k_{13}} \cdot \mathbf{r_{13}} + \mathbf{k_{23}} \cdot \mathbf{r_{23}})} \\
    & \times \delta_\mathrm{D}(\mathbf{k_{12}}+\mathbf{k_{13}}+\mathbf{k_{23}}) 
    \end{split}
\end{equation}
where we have introduced the bispectrum, defined as 

\begin{equation}
    \label{eq:bisp_fund}
     \left< \delta_m(\mathbf{k_1}) \delta_m(\mathbf{k_2}) \delta_m(\mathbf{k_3}) \right> = (2\pi)^3 \delta_\mathrm{D}(\mathbf{k_{12}}+\mathbf{k_{13}}+\mathbf{k_{23}}) B_m(k_1, k_2, k_3).
\end{equation}
The formal definition in Eq \eqref{eq:bk_transform} is of no practical use because the corresponding transformation requires performing integration in a 6-D space. The complexity of the problem can be reduced by integrating the angular variables and by using the expansion of a plane wave in spherical harmonics (see Appendix in \citep{Slepian2015a} for a usage)

\begin{equation}
    e^{i \bf{k} \cdot \bf{r}} = 4 \pi \sum_{l = 0}^{+\infty} \sum_{m = -l}^{l} i^l j_l(kr) Y_{lm}(\hat{\bf{r}})Y^*_{lm}(\hat{\bf{k}} ),
\end{equation}
so that Eq. \ref{eq:bk_transform} reduces to

\begin{equation}
    \label{eq:zeta_from_bkell}
    \zeta_{\ell}(r_{12}, r_{13}) = (-1)^{\ell} \int \frac{k_{12}^2 k_{13}^2 \mathrm{d}k_{12} \mathrm{d}k_{13}}{(2 \pi^2)^2} B_{\ell}(k_{12}, k_{13}) j_{\ell}(k_{12} r_{12}) j_{\ell}(k_{13} r_{13}).
\end{equation}
The integration in Eq. \ref{eq:zeta_from_bkell} will be the cornerstone of this work to obtain a 3PCF model at higher perturbative order compared to the tree level. The procedure we used to model 3PCF from bispectrum is summarized in the block diagram in Fig. \ref{fig:iteration_summary}. In step 1, we  model the matter bispectrum. In step 2, we compute its multipoles through the Gauss-Legendre quadrature method. Step 3 evaluates the Hankel transform \ref{eq:zeta_from_bkell} using the 2D-FFTLog algorithm \cite{Fang2020}. Finally, these multipoles are summed over to obtain the 3PCF model.

\begin{center}
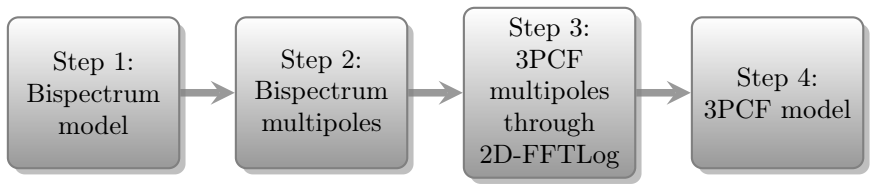

\smartdiagramset{uniform color list=white!60!black for 6 items,
    back arrow disabled=true, module minimum width=2cm,
    module minimum height=2cm,
    module x sep=3cm,
    text width=2cm,
    additions={
        additional item offset=0.5cm,
        additional item width=2cm,
        additional item height=2cm,
        additional item text width=3cm
      }
}
\smartdiagram[flow diagram:horizontal]{Step 1: Bispectrum model, Step 2: Bispectrum multipoles, Step 3: 3PCF multipoles through 2D-FFTLog, Step 4: 3PCF model} 
\captionof{figure}{Flow chart that summarizes the procedure used to generate the 3PCF model from the bispectrum one.}
\label{fig:iteration_summary}
\end{center}

In Step 1, we consider both the leading (LO-2D) and two different next-to-leading order (NLO-SPT and NLO-EFT) models for the matter bispectrum \cite{Bernardeau2002}, that will be described in detail in the next section. Then we expand the models in multipoles in step 2, using the Gauss-Legendre quadrature method. The different bispectrum models used in this work are listed in Table \ref{tab:models_resume} along with the technique used to perform the Hankel transform coming from the multipoles expansion of \ref{eq:bk_transform} \MG{and the degrees of freedom involved in the considered models}. The LO-1D model is the leading order 3PCF model proposed by \citep{SlepianEisenstein2017} obtained by transforming the leading order bispectrum model (see \citep{Bernardeau2002} for a review), using 1D-FFTLog thanks to analytical simplifications - limited to the leading order case - by integrating over the angular part of Eq. \ref{eq:bk_transform}. The LO-2D is equivalent to the LO-1D model, except that the 3PCF model is obtained by integrating a 2D integral using 2D-FFTLog. The NLO-SPT model is based on the one-loop bispectrum (see \citep{Scoccimarro98, Bernardeau2002} for a review) and uses the 2D-FFTLog to evaluate the analogous next-to-leading order SPT 3PCF model. The NLO-EFT model implements the bispectrum model in the Effective Field Theory picture \citep{Bauman, Baldauf2015, Angulo:2014tfa} and again generates the corresponding 3PCF model through the 2D-FFTLog transform. 
Unlike the other cases, the NLO-EFT model relies on four free parameters that are evaluated by fitting the data.

\begin{table}[H]
    \begin{center}
    \renewcommand{\arraystretch}{1.4}
    \begin{tabular}{c|c|c}
         \hline
         3PCF model name & Method & Degrees of freedom \\
         \hline
         \ LO-1D & 1D-FFTLog & 0 \\  
         \ LO-2D & 2D-FFTLog & 0\\
         \ NLO-SPT & 2D-FFTLog & 0\\
         \ NLO-EFT & 2D-FFTLog & 4\\
        \hline
    \end{tabular}
    \end{center}
    \caption{Summary of models used in this paper, methods used to compute them and their degrees of freedom.
    \label{tab:models_resume}
    }
\end{table}

\subsection{Perturbative expansion}
\label{subsec:SPT}

The formation and the evolution of cosmic structures are governed by collisionless dark matter that can be modelled as a self-gravitating fluid obeying the Vlasov equation in an expanding background.
As an assumption, SPT sets to zero the second velocity moment of the phase-space distribution function and assumes irrotational flow, allowing us to 
derive, Euler and Poisson equations for the matter density contrast and divergence of peculiar velocities. Their solution can be obtained perturbatively by expanding the density contrast in terms of the linear solution $\delta^{n}_{L}(\mathbf{k})$ and linearised transport equation solution $\delta^{1}(\mathbf{k}, z) = D(z) \delta_{L}(\mathbf{k})$, where $D(z)$ is the linear growth factor. For an Einstein-de Sitter (EdS) cosmology and given appropriate kernels of gravitational coupling between Fourier modes, one obtains:

\begin{equation}
\label{eq:B_tree}
    \delta(\mathbf{k}, z) = \sum_{n=1}^{\infty} D(z)^{n} \delta^{n}(\mathbf{k})
\end{equation}
with 
\begin{equation}
    \delta^n(\mathbf{k}) = \int \frac{\mathrm{d}^3 \mathbf{k}_1 ... \mathrm{d}^3 \mathbf{k}_n}{(2\pi)^{3(n-1)}} \delta_D(\mathbf{k} - \mathbf{k}_{1..n}) F_{n}(\mathbf{k}_1, ..., \mathbf{k}_n)  \delta_L(\mathbf{k}_1)...\delta_L(\mathbf{k}_n) 
\end{equation}
where we report the recursive relation and the expressions
for $F_{n}(\mathbf{k}_1, ..., \mathbf{k}_n)$ in Appendix \ref{AppendixA}. 
\subsubsection{Modelling bispectrum}
Under the assumption of Gaussian random field $\delta_L(\mathbf{k})$, the bispectrum given in Eq. \ref{eq:bisp_fund} can be modelled at the so-called tree-level

\begin{equation}
    B_{\mathrm{SPT}}^{\mathrm{tree}}(\mathbf{k}_1, \mathbf{k}_2, \mathbf{k}_3, z) = 2F_2(\mathbf{k}_1, \mathbf{k}_2) P_\mathrm{L}(k_1, z) P_\mathrm{L}(k_2, z) + \mathrm{cyc}.   
\end{equation}
where the matter density contrast $\delta(\mathbf{k}, z)$ is expanded at second order. 

So far, the existing matter 3PCF model is based on leading-order perturbative expansion in Standard Perturbation Theory, which allows one to rely on 1D-FFTLog.
Our goal is to go beyond the first order. Therefore we consider the next-to-leading order bispectrum. This means that we  
compute the standard four integrals to evaluate the one-loop bispectrum expansion \citep{Scoccimarro98, Bernardeau2002}
\begin{equation}
\begin{split}
    & B^\mathrm{one-loop}_{\mathrm{SPT}}(k_1, k_2, k_3, z) = B_{222}(k_1, k_2, k_3, z) + 
     B_{321-I}(k_1, k_2, k_3, z) \\&  + B_{321-II}(k_1, k_2, k_3, z) + B_{411}(k_1, k_2, k_3, z). 
    \label{eq:BSPT}
\end{split}
\end{equation}

\subsubsection{Infrared resummation}
\label{subsec:IR}

To improve the accuracy of the model at the scale of the baryonic acoustic oscillations (BAO) feature, it is a common practice to perform the so-called infrared (IR) resummation \citep{Blas, Baldauf2015, Ivanov}.
This is a crucial step to correctly account for higher perturbative order contributions, as best shown  by the case of the 2-point correlation function in configuration space. In that case, neglecting resummation would generate a spurious double peak around the BAO scale, making it impossible to use it as a standard ruler for precise observational tests.
We also include the IR resummation in our models to avoid generating similar spurious features in the 3PCF. To include the IR resummation, we first decompose the power spectrum into a smooth and wiggly component so that \citep{Seo_2008}
\begin{equation}
    P(k) = P_{\mathrm{nw}}(k) + P_{\mathrm{w}}(k). 
\end{equation}
The effect is splitting the oscillating contribution from BAO imprinting into the power spectrum from the smoothed part from the remaining  non-baryonic physics ruling the shape of the power spectrum. Then, we damp the wiggling part to recover the infrared tree-level power spectrum \citep{Ivanov}
\begin{equation}
    P^{\mathrm{IR}}(k)= P_{\mathrm{nw}}(k) + e^{-k^2 \Sigma^2}P_{\mathrm{w}}(k), 
\end{equation}
where the damping factor is given by the relative displacement two-point function in the Zel'dovich approximation at the BAO scale \citep{Eisenstein_2007}
\begin{equation}
    \Sigma^2 = \int_0^{k_\mathrm{S}} \frac{\mathrm{d^3q}}{(2\pi)^3}\frac{P_\mathrm{nw}(q)}{3q^2} \left [ 1-j_0(\frac{q}{k_\mathrm{BAO}}) + 2j_2(\frac{q}{k_\mathrm{BAO}}) \right],
\end{equation}
where $j_n(x)$ are spherical Bessel functions and $k_\mathrm{BAO} = \pi/\ell_\mathrm{BAO}$ with $\ell_\mathrm{BAO} = 110 \mathrm{Mpc}/h$. The $k_\mathrm{S}$ cutoff is commonly fixed to $k_\mathrm{S} = 0.2 h/\mathrm{Mpc}$ \citep{Sanchez_2016, Blas} and we use this value.
To use the IR procedure at the bispectrum level, we compute bispectrum contributions using the $P^\mathrm{IR}(k)$ (both tree-level and one-loop computations), and then we compute multipoles. This allows us to use the leading and next-to-leading orders, connected with tree-level and one-loop contributions by the infrared resummation scheme 

\begin{align}
   \label{eq:SPT_LO_Bk}
   &  B_{\mathrm{LO}}(k_1, k_2, k_3) = B_{\mathrm{tree}}^{\mathrm{IR}}(k_1, k_2, k_3), \\
   \label{eq:SPT_NLO_Bk}
   &  B_{\mathrm{NLO}}(k_1, k_2, k_3) = B_{\mathrm{tree}}^{\mathrm{IR}}(k_1, k_2, k_3) + B_{\mathrm{one-loop}}^{\mathrm{IR}}(k_1, k_2, k_3). 
\end{align}

\subsubsection{Effective field theory based models}
\label{subsec:EFT}

Besides the Standard Perturbation Theory bispectrum model, we also consider the second-order expansion bispectrum  model obtained from the Effective Field Theory (EFT) \citep{Carrasco2014, Hertzberg2015, Angulo:2014tfa, Baldauf2015}. The purpose of EFT is to provide an effective description of long-wavelength 
modes where SPT usually fails. The EFT procedure uses an effective stress tensor expressed in terms of all operators of long-wave-length density and velocity fields. Taylor expanding the effective stress tensor gives us an infinite series of unconstrained parameters associated with each perturbative expansion. These parameters can be treated as coupling constants in the Wilson approach to renormalisation \citep{Wilson}.

EFT coupling coefficients can be seen as counterterms that can be used to cancel the dependence of SPT on the UV scale and to model a non-ideal stress tensor. The remaining cutoff-independent part of the counterterms quantifies the impact of physics that cannot be described perturbatively by introducing effective interaction on long-wavelength modes. It is possible to express EFT contribution at second order as the sum of 4 counterterms \citep{Eggemeier2019, Alkhanishvili:2021pvy, Garny2022}:
\begin{equation}
\begin{split}
    B_{\mathrm{EFT}}(k_1, k_2, k_3)  = B_{\mathrm{NLO}}(k_1, k_2, k_3) + & B_{c_0} + B_{c_1} + B_{c_2} + B_{c_3}. 
\end{split}
\end{equation}
where

\begin{align}
    \label{eq:EFT_1}
    B_{c_0} =  c_0(z) [D(z)]^4 [2P_\mathrm{L}(k_1)P_\mathrm{L}(k_2)\tilde{F}^{(s)}_2(\mathbf{k}_1, \mathbf{k}) + 2 \ \mathrm{perms} \\
     -2k_1^2P_\mathrm{L}(k_1)P_\mathrm{L}(k_2)F_2(\mathbf{k}_1, \mathbf{k}) + 5 \ \mathrm{perms} ], \notag \\
    B_{c_1} = -2c_1(z) [D(z)]^4 k_1^2 P_\mathrm{L}(k_2)P_\mathrm{L}(k_3) + 2 \ \mathrm{perms}, \\
    B_{c_2} = -2c_2(z) [D(z)]^4 k_1^2 \frac{(\mathbf{k}_2 \cdot \mathbf{k}_3)^2}{k_2^2 k_3^2}P_\mathrm{L}(k_2)P_\mathrm{L}(k_3) + 2 \ \mathrm{perms},  \\
    \label{eq:EFT_2}
    B_{c_3} = -2c_3(z) [D(z)]^4 (\mathbf{k}_2 \cdot \mathbf{k}_3) [\frac{\mathbf{k}_1 \cdot \mathbf{k}_2}{k_2^2}+ \frac{\mathbf{k}_1 \cdot \mathbf{k}_3}{k_3^2}]P_\mathrm{L}(k_2)P_\mathrm{L}(k_3) + 2 \ \mathrm{perms},   
\end{align}
where $\tilde{F}^{(s)}$ is defined in Appendix \ref{AppendixA}. 
EFT can  be used, in the end,  as an extension of SPT in which extra degrees of freedom are involved to take account of the UV pathological behaviour of loop integrals.  

\subsection{2D-FFTLog}
\label{Method_Motivation_2DFFTLog}

A common problem in physics is the numerical evaluation of integrals involving the product of two or more Bessel functions as in Eq. \ref{eq:zeta_from_bkell}.
Evaluating these integrals represents the main computational burden of our 3PCF modelling procedure.
Evaluating the integral from a grid $N_r \times N_r$ using the standard quadrature method to compute Eq \eqref{eq:zeta_from_bkell} requires an order of $N_r^2 N_k^2$ steps where $N_r$ and $N_k$ are the number of $k$ and $r$ points sampled. 
Furthermore, the rapid and oscillatory behaviour of Bessel functions requires 
performing many integration steps, making the accurate evaluation of the integrals computationally challenging. 
The 1D-FFTLog method, originally conceived to address atomic physics problems (\cite{Talman1978}) and then applied to cosmology by \citep{Hamilton:1999uv}, has been used over the years to efficiently evaluate Fourier transform with logarithmic variables involving single Bessel integration of the form $\int_0^\infty \mathrm{d}k f(k) j_\ell(kr)$ where $f(k)$ is a generic smooth function. The main idea of the 1D-FFTLog algorithm is using the expansion $f(k) = \sum_m c_m k^{z_m}$ with $z_m$, in general, a complex number to obtain a term that can be integrated analytically to
speed up the computation by evaluating a sum over  $c_m$ Fourier coefficients instead of a 1D-integral. The FFTLog approach has proven to be useful, as can be seen by three notable and quite different applications of the method.
The first one is the evaluation of the Bessel integrals in the angular power spectra and bispectra expressions \citep{Assassi, Grasshorn, Schoneberg}. The second one by \cite{Slepian:2019} 
consists of using 1D-FFTLog to evaluate Bessel integrals coming from the deconvolution of multi-dimensional integrations. Finally, FFTLog  has been used to model higher-order statistics in Fourier space. 
In this latter case, there is no Bessel integral involved. Instead, the idea is to obtain integrable expressions to which the FFTLog tool can be applied. Examples include 
one-loop \citep{Schmittfull:2016A, McEwen:2016} and two-loop \citep{Schmittfull:2016B, Simonovic:2017mhp, Slepian_2018} perturbation models. \MG{Furthermore, an application of the 2D-FFTLog algorithm to the tree-level anisotropic 3PCF can be found in \cite{Umeh2021}.} 


For the 3PCF model, in general, one has to consider the  2D-dimensional integral in Eq. \ref{eq:zeta_from_bkell}, which involves Bessel functions. In this case, the usual 1D-FFTLog approach is of no use since one cannot separate terms involving $k_{12}$ and $k_{13}$. 
\MG{The 2D-FFTLog method, recently proposed in \cite{Fang2020}, relies on the same idea as the 1D-FFTLog, but uses a double power law expansion in order to accurately evaluate integrals involving the product of two Bessel functions}. This work was originally addressed to the computation of real-space 3PCF covariances, similar to Eq. \ref{eq:zeta_from_bkell}. 
The 2D extension  of the FFTLog algorithm is sensitive, as the 1D version, to all sources of aliasing and ringing \citep{Hamilton:1999uv}. The grid we use is 2D, parametrised by two $k_1$, $k_2$ after integrating over the angle between the two, $\theta$. In general, grid spacing $\Delta k$, regulates the integration accuracy. The choice of $k_{\mathrm{min}}$ and $k_\mathrm{max}$ is crucial to avoid aliasing and ringing effects. Zero padding is also advised to reduce the possibility of generating spurious wiggles. \MG{As with any FFT-based algorithm, the acceleration method is based on biasing integrands by power law weights.
In the implementation of the 2D-FFTLog 
the quantity to be transformed is the dimensionless bispectrum multipoles $\Delta_\ell(k_{1}, k_{2})$ (see Appendix \ref{AppendixC}). Being $k_1$ and $k_2$ discrete and logarithmically sampled arrays --- whose elements denoted by the $i$th index for $i = 0, 1, ..., N-1$ --- this function can be expanded as}
\begin{equation}
\begin{split}
    \Delta_\ell(k_{p}, k_{q}) = 
    \frac{1}{N^2} \sum_{m = -\frac{N}{2}}^{N/2} \sum_{n = -\frac{N}{2}}^{N/2} \tilde{c}_{\ell, mn} k_0^{-i\eta_m} k_0^{-i\eta_n} {k_{p}^{\nu_1 + i\eta_m}} {k_{q}^{\nu_2 + i\eta_n}},
    \label{eq:delta_fftlog}
\end{split}
\end{equation}
\MG{where $\eta_m = 2\pi \frac{m}{N\Delta_{ln\textit{k}}}$, $N$ is the size of $k$-array, $\nu_1$, $\nu_2$ are the so-called \textit{bias parameters}, i.e. the real part of the power laws, $\Delta_{\mathrm{ln}\textit{k}}$ is the linear spacing in $\ln k$ so that $k_q = k_0 \ \mathrm{exp}(q \Delta_{\mathrm{ln}\textit{k}})$. The coefficients $c_{\ell, mn}$ are given by the discrete Fourier transform as follows} 

\begin{equation}
    \tilde{c}_{\ell, mn} = \sum_{p = 0}^{N-1} \sum_{q = 0}^{N-1} \frac{\Delta_\ell(k_{p}, k_{q})}{k_{p}^{\nu_{1}} k_{q}^{\nu_{1}}} \mathrm{e}^{-2\pi i (mp + nq)/N}. 
    \label{eq:clm_fftlog}
\end{equation}
\MG{This power law expansion facilitates the Fourier transform.
 To remove sharp edges at the boundary of $c_{\ell, mn}$ we filter the expansion coefficients as follows}

\begin{equation}
    c_{\ell, mn} = W_{m}W_{n}\tilde{c}_{\ell, mn}  \,
    \label{eq:clm_fftlog_filt}
\end{equation}
\MG{where \textbf{W} is a one-dimensional window function, whose $m-$th and $n-$th elements are represented by $W_m$, $W_n$ (for details, see \cite{McEwen:2016}). For details and the application to the 3PCF case see Appendix \ref{AppendixC}.}

In our application of the 2D-FFTLog, we set the bias parameters $v_1 = 1.01$ and $v_2 = 1.01$, equal to the optimal choice determined in \citep{Fang2020}.
By applying zero-padding, we use $N_{\mathrm{pad}} = 200$ on the small and large wavevectors' sides to reduce the ringing effect. The dimension of the grid is fixed to $256 \times 256 \times 51$, since we sample 256 wavenumbers along the 
 $k_1$ and $k_2$ directions and consider 51 angle values sampled using the Gauss-Legendre quadrature method, i.e. the same method we use to compute bispectrum multipoles we use as an input to the 2D-FFTLog algorithm. 
 Finally, we use $k_{\mathrm{min}} = 3 \times 10^{-4}  \ h \ \mathrm{Mpc}^{-1}$ and $k_{\mathrm{max}} = 5 \ h \ \mathrm{Mpc}^{-1}$, to  avoid aliasing and to damp nonlinear contributions to the 
 power spectra from scales in which the perturbative approach fails. The aforementioned choice has been tested and validated in Appendix \ref{AppendixB}.

\subsection{3PCF models}
\label{subsec:3PCF_models}
The models we are considering in this paper are listed in Table \ref{tab:models_resume}. We mainly use the same strategy to get access to 3PCF modelling, starting from Fourier Space and then inverse-Fourier transforming to have 3PCFs but using the two different methods introduced in the previous Section. 

First, we generate the LO-1D model following the same procedure as \citep{SlepianEisenstein2017} and summarised in Appendix \ref{AppendixD}. 
The LO-1D model is well-established, equivalent to the  3PCF models of \citep{Jing2003} and \citep{Barriga2001}. The Fourier-space analogue is the leading order bispectrum model (see \citep{Scoccimarro98} and \cite{Bernardeau2002} for a review).
We stress that Eqs. \ref{eq:xi_SE_1} and \ref{eq:xi_SE_2} that define the model are one-dimensional integrals, including the Bessel functions that can be computed through the 1D-FFTLog algorithm. We will use the LO-1D model as a benchmark to validate the implementation of the 2D-FFTLog used for other models. 

Secondly, 
we 
use 2D-FFTLog to generate LO-2D 3PCF model from the bispectrum model
Eq. \ref{eq:SPT_LO_Bk}. This is equivalent to the LO-1D case, except that we use 2D-FFTLog in the process.
Then 2D-FFTLog is also used to generate the NLO-SPT 3PCF model from the bispectrum model Eq. \ref{eq:SPT_NLO_Bk} and the NLO-EFT 3PCF model from the bispectrum model defined by the set of 
 Eqs. \ref{eq:EFT_1}-\ref{eq:EFT_2}. 
 
 In all cases, the input bispectrum multipoles are computed using the Gauss-Legendre quadrature method 
 to evaluate the integral in Eq. \ref{eq:leg_transf}.
 
To compare with measured quantities, we provide bin-averaged predictions, i.e.
for every pairs $(r_{12},r_{13})$, we compute

\begin{equation}
        \label{eq:zeta_binave}
        \zeta_\ell (\bar{r}_{i}, \bar{r}_{j}) = \frac{\int_{\bar{r}_{i, \mathrm{min}}}^{\bar{r}_{i, \mathrm{max}}} dr_1 \int_{\bar{r}_{i, \mathrm{min}}}^{\bar{r}_{i, \mathrm{max}}} dr_2 r_1^2 r_2^2 \zeta_\ell(r_1, r_2)}{\int_{\bar{r}_{i, \mathrm{min}}}^{\bar{r}_{i, \mathrm{max}}} dr_1 r_1^2 \int_{\bar{r}_{i, \mathrm{min}}}^{\bar{r}_{i, \mathrm{max}}} dr_2 r_2^2}, 
\end{equation}
where $\bar{r}_{i}$ and $\bar{r}_{j}$ are the i-th and j-th bins with width equal to $\Delta_\mathrm{{bin}} = (r_{i, j, \mathrm{max}} - r_{i, j, \mathrm{min}})/2$. In practice, we do not use Eq. \ref{eq:zeta_binave} to average the models on the bin size. Instead, we directly implement the binned average on the FFTLog, as presented in \citep{Fang2020}. For details, see Appendix \ref{AppendixC}. 

Finally, to represent the 3PCF models we use the following ordering convention  $r_{12} \leq r_{13} \leq r_{23}$. 

\section{The DEMNUni simulations}
\label{sec:measure}

\textcolor{black}{In this work, 3PCF measurements are performed on the ``Dark Energy and Massive Neutrinos Universe'' (DEMNUni) N-body simulations~\cite{DEMNUni_simulations}. The DEMNUni simulations have been produced with the aim of investigating large-scale structures in the presence of massive neutrinos and dynamical dark energy, and they were conceived for the nonlinear analysis and modelling of different probes, including dark matter, halo, and galaxy clustering~\cite{Castorina_2015,Moresco2017,Zennaro2018,Ruggeri2018,Bel2019,Parimbelli2021,Parimbelli2022, Baratta_2022, SylvainBG2023, SHAM-Carella_in_prep}, weak lensing, CMB lensing, SZ and ISW effects~\cite{Roncarelli2015,DEMNUni_simulations,fabbian2018, Beatriz_2023}, cosmic void statistics~\cite{Kreisch2019,Schuster2019,Verza2019,Verza_2022a,Verza_2022b}, and cross-correlations among these probes~\cite{Vielzeuf2022_inprep,Cuozzo2022_inprep}.
To this end, they combine a good mass resolution with a large volume to include perturbations both at large and small scales. In fact, these simulations follow the evolution of 2048$^3$ cold dark matter (CDM) and, when present, 2048$^3$ neutrino particles in a box of side $L=2 \ {\rm Gpc}/h$. The fundamental frequency of the comoving particle snapshot is, therefore, $k_{\rm F} \approx 3\times 10^{-3} \ h/$Mpc, while the chosen softening length is 20 kpc/$h$. The simulations are initialized at $z_{\rm ini}=99$ with Zel'dovich initial conditions. The initial power spectrum is rescaled to the initial redshift via the rescaling method developed in Ref.~\cite{rescaling_IC-Zennaro+17}. Initial conditions are then generated with a modified version of the \texttt{N-GenIC} software, assuming Rayleigh random amplitudes and uniform random phases.  The simulations were performed using the tree particle mesh-smoothed particle hydrodynamics (TreePM-SPH) code GADGET-3, an improved version of the code described in Ref.~\cite{Springel2005}, specifically modified in Ref.~\cite{Viel2010} to account for the presence of massive neutrinos. These are assumed to come in three mass-degenerate species. The sum of their masses is varied over the values $M_\nu=0, 0.16, 0.32$ eV. For each value of $M_\nu$, five different simulations were run: one with ($w_0=-1$, $w_a=0$) and four with various combinations of $w_0=(-0.9, -1.1)$ and $w_a=(-0.3, 0.3)$, for a total of 15 simulations. For each simulation, 63 snapshots, logarithmically equispaced in the scale factor $a$, are saved. About 400 TB of data in particle comoving snapshots, halo and galaxy catalogues, projected density maps, and power spectra of different particle species are stored and available upon request.}

In this work, we are interested in the spatial distribution of CDM particles and in the standard $\Lambda$CDM case alone.

Since the number of particles is too large for an efficient estimate of their 3PCF, we down-sample the population to $278^3$ particles. 
\textcolor{black}{The dilution factor $2.5 \permill$ used here represents a reasonable tradeoff between the computational cost of the 3PC estimator and the shot-noise error}. 

The cosmological parameters of the $\Lambda$CDM model used in the simulation are similar to the \textit{Planck} 2013 \citep{Planck2014} ones:
 $\Omega_b = 0.05$, $\Omega_M = 0.32$, $h=0.67$, $n_s=0.96$ and $\sigma_8 = 0.846$. This is the parameter set that will be used throughout this work
 
Finally, we will only use the $z= 0.49$ and $z= 1.05$ snapshots since they are in the redshift range in which 3PCF measurements from large datasets are and will be available, $[0.7-2]$ for Euclid \cite{EuclidReport}, $[0.6-1.7]$ for DESI \cite{DESI2016} and $[1.1-2.6]$ for WFIRST \cite{Eifler_2021}. Also, the Gaussian errors indicated in the plot account for the cosmic variance and the shot noise error for a sample of objects comparable, in size and volume, to those of currently available surveys (BOSS at $z=0.49$) or future datasets at $z=1.05$ (DESI, Euclid and WFIRST).

\subsection{Dark matter clustering measurements}
\label{sec:clustering_meas}

For an efficient triplet counting procedure, we use the Szapudi-Szalay estimator complemented with the Spherical Harmonics Decomposition (SHD) technique introduced  \citep{Slepian_2015, Veropalumbo2021}. SHD 
significantly reduces the computational cost of direct triplet counting, 
since the scaling with the number of objects is proportional to 
$N^2$ rather than  $N^3$. 
We estimate $\hat{\zeta}_{\ell}(r_{12}, r_{13})$, up to $\ell_{max}=10$ and then we obtain the approximate estimate if the  triangle-binned 3PCF:
\begin{equation}
    \hat{\zeta}(r_{12}, r_{13}, r_{23}) = \sum_{\ell = 0}^{\ell_{max}} \hat{\zeta}_{\ell}(r_{12}, r_{13}) \widetilde{\mathcal{P}}_\ell(r_{12}, r_{13}, r_{23}); 
\end{equation}
where $\widetilde{\mathcal{P}}_\ell(r_{12}, r_{13}, r_{23})$ is the Legendre polynomial of order $\ell$ integrated over the triangle \citep[see Appendix A of][]{Veropalumbo2021}. This estimator is known to be \MG{difficult in handling the isosceles triangle configurations} ($r_{12}=r_{13}$) using  a reasonably small number of multipoles $\ell_{max}$ \citep{Veropalumbo2021}. 

This problem also extends to nearly isosceles triangles for which $r_{13} \simeq r_{12}$. For this reason, and to separate these cases from the results obtained for all other configurations, 
we characterise measurements using two parameters the minimum separation, $r_{\mathrm{min}}$, 
and the minimum value, $\eta_{\mathrm{min}}$, for the relative difference
\begin{equation}
    \eta \equiv \frac{r_{13} - r_{12}}{\Delta r}\,.
    \label{eq:eta}
\end{equation}
For a given set $\Delta r$ - i.e. the bin-width -, $r_{\mathrm{min}}$ and $r_{\mathrm{max}}$ the 
case $\eta_{\mathrm{min}}=0$ includes all triangles. The case $\eta_{\mathrm{min}}=1$ includes all triangles with 
$|r_{13}-r_{12}| > \Delta r$. And so on. Inputs to the 3PCF estimator are the spatial distribution of the objects and 
the spatial distribution of a \textit{random} set of unclustered objects distributed in the same volume and with the same selection function as the data. 
Moreover, since we measure the 3PCF in a computational simulation of an N-body experiment, the \textcolor{black}{triplet counter algorithm accounts for the periodicity of the box.}
We use a random dataset 20 times larger than the real dataset to reduce the noise \MG{associated to the shot-noise of the random sample.}
\textcolor{black}{The random sample is needed to consistently subtract the disconnected part of \MG{the probability of finding a triplet of clustering objects, i.e.} the 3PCF.} To reduce the computational \textcolor{black}{effort}, we used the random splitting technique \citep{elina19, Slepian_2015}; \textcolor{black}{we repeat the 3PCF measurement of the same sample $20$ time with a different random catalogue the same size of the data, and then we average the estimates. This also allows us to explicitly determine the \MG{single contribution on the total} error coming from this procedure}. 

\subsection{Gaussian covariance}
\label{subsec:cov}
The final ingredient to compare models with simulated data is the 3PCF covariance matrix $\mathbb{C}$. Since a numerical estimate of $\mathbb{C}$ is too computationally demanding, we use the Gaussian model of \citep{Szapudi2001}  to obtain a theoretical expression for the 3PCF covariance matrix $\mathbb{T}$. Thanks to the periodic boundary condition, we can further simplify the model expression and ignore mode coupling due to the sample geometry. \textcolor{black}{We also consider the effect of binning in the computation.}
With these assumptions, we can obtain a simplified expression for the Gaussian 3PCF covariance matrix for each Legendre coefficient, $T_{\ell, \ell^\prime}(r_{12}, r_{13} ; r_{12}^\prime, r_{13}^\prime)$ and use it to form the triangle-binned 3PCF covariance matrix:
\begin{align}
    \label{eq:cov_resum}
    T(r_{12}, r_{13}, r_{23} ; r_{12}^\prime, r_{13}^\prime, r_{23}^\prime )  =  \sum_{\ell, \ell^{\prime}=0}^{\ell_{max}} 
    T_{\ell, \ell^\prime}(r_{12}, r_{13} ; r_{12}^\prime, r_{13}^\prime) & \\  \mathcal{P}_{\ell}(\mu) \mathcal{P}_{\ell^{\prime}}(\mu^\prime) \notag,
\end{align}
where $\mathcal{P}_{\ell}$ are the Legendre polynomials of degree $\ell$ and $\mu$ is the cosine angle between $r_{12}$ and $r_{13}$.
The expression for $T_{\ell, \ell^\prime}(r_{12}, r_{13} ; r_{12}^\prime, r_{13}^\prime)$ and the details of each estimate can be found in \cite{Slepian_2015}.
Its value is contributed by cosmic variance, which in turn depends on the volume of the sample $V$ and on the matter power spectrum $P(k)$, and by the shot noise of the discrete tracers with mean density $\bar{n}$. 
The values of  $V$ and $\bar{n}$ are taken from the simulation box whereas 
we use one-loop SPT to model the power spectrum.

The use of the splitting method to estimate the 3PCF introduces an additional error, $\sigma_{r}$, on top of the shot noise one. We model its contribution as an additional diagonal term to the theoretical covariance, i.e.
\begin{equation}
    \mathbb{C} = \mathbb{T} + \mathbf{\sigma_{r}}^2 \mathbb{I} \, 
    \label{eq:cov}
\end{equation}
\GM{being $\sigma_{\mathbf{r}}$ computed throughout the scatter from different 3PCF estimates using different randoms} and we use $\mathbb{C}$ in the $\chi^2$ analysis.

The accuracy of the Gaussian approximation for the 3PCF covariance matrix  has been assessed by \cite{Slepian_2015, Veropalumbo2022}. Based on their results, we expect that the Gaussian model provides an unbiased estimate of the covariance matrix for separations $r\geq40 \, h^{-1} \mathrm{Mpc}$ whereas it underestimates the uncertainties and their correlation on smaller scales. To correctly evaluate the error budget, we renormalise the covariance matrix $\mathbb{C}$ by the factor $\alpha = 1.4$ to get a meaningful value of $\chi^2$ around unity at sufficient large separations for the LO model
\begin{equation}
    \mathbb{C}' = \frac{\mathbb{C}}{\alpha}.
    \label{eq:cov_2}
\end{equation}
We will use $\mathbb{C}'$ throughout the rest of the paper. The impact of the treatment of uncertainties will be discussed in the framework of our $\chi^2$ analysis, see Sec. \ref{sec:results}.

\section{Comparing models to data}
\label{sec:matching}

To assess the goodness of the model, we perform a $\chi^2$ comparison with the data. We estimate
\begin{equation}
    \chi^2 = \sum_{i, j} (M_i - D_i)C^{-1}_{ij}  (M_j - D_j)
\end{equation}
where $M_i$ is the model vector, $D_i$ is the data vector and $C_{ij}$ is the Gaussian covariance described in the previous section. 
The Gaussian hypothesis may fail on scales as small as those considered in our analysis.
However, any inaccuracy introduced in the error estimate will affect all 
$\chi^2$ analyses in a similar fashion and, therefore, will not bias the comparison between models.

We estimate the $\chi^2 $ statistics in the 
eigenspace  i.e.
\begin{equation}
    \chi^2 = \sum_{i} \frac{(\Lambda_{ij}M_j - \Lambda_{ij}D_j)^2}{\lambda_i}
\end{equation}
where $\Lambda_{ij}$ are the coefficients of the eigen-matrix \MG{of the Gaussian covariance}, $\lambda_i$ are the positive eigenvalues, \MG{thus the data and models being decomposed in the eigen-basis of the covariance}. In our analysis, we  use the reduced chi-square $\chi^2_r = \chi^2/\nu$, where $\nu$ indicates the number of degrees of freedom. 

To quantify the relative performance of nested models such as NLO-SPT and NLO-EFT we use the cumulative chi-square difference normalised by the degree of freedom of the NLO-SPT (i.e. the number of triangle configurations given the set $\Delta r$, $r_{\mathrm{min}}$ and $\eta$)

\begin{equation}
    \frac{\Delta \chi^2}{\nu_{\mathrm{NLO-SPT}}} = \frac{\chi^2_{\mathrm{NLO-SPT}} - \chi^2_{\mathrm{NLO-EFT}}}{\nu_{\mathrm{NLO-SPT}}}.
    \label{eq:diff_chi}
\end{equation}

We will also consider the residuals between models and data normalised to the statistical errors:
\begin{equation}
\label{eq:res}
R_{i} = \frac{M_i - D_i}{\sigma_i}
\end{equation}
where $\sigma_i$ are the errors extracted from the diagonal elements of $C_{ij} $. 
The level of agreement between models and data is expected to vary with scale. To assess the relative goodness of the LO and NLO models and how it changes with scale we compute the average residuals on all triangle configurations and some selected ones, and compare them in the summary statistics 
\begin{equation}
    \left< \Delta R \right> = \frac{1}{N} \sum_i^{N} |R_i^{\mathrm{NLO}}| - |R_i^{\mathrm{LO}}|
    \label{eq:delta_res},
\end{equation}
where $N$ is the number of configurations examined.

\section{Results}
\label{sec:results}
We now compare the different 3PCF models introduced in Section \ref{sec:models} with the measurements performed on the DEMNUni simulations. In doing so, we distinguish, somewhat arbitrarily, between large ($ > 40 h^{-1} \mathrm{Mpc}$) and small ($ < 40 h^{-1} \mathrm{Mpc}$) separations to indicate the scales on which leading and next-to-leading 3PCF models are expected to provide different and matching predictions, respectively.
In addition, we will focus on triplets in which at least one of the sides matches the BAO scale of $\simeq 110 h^{-1}\mathrm{Mpc}$. We call these the BAO configurations.

\subsection{Comparisons among 3PCF models}
\label{subsec:models}
\begin{figure}[H]
\includegraphics[width=1.0 \textwidth]{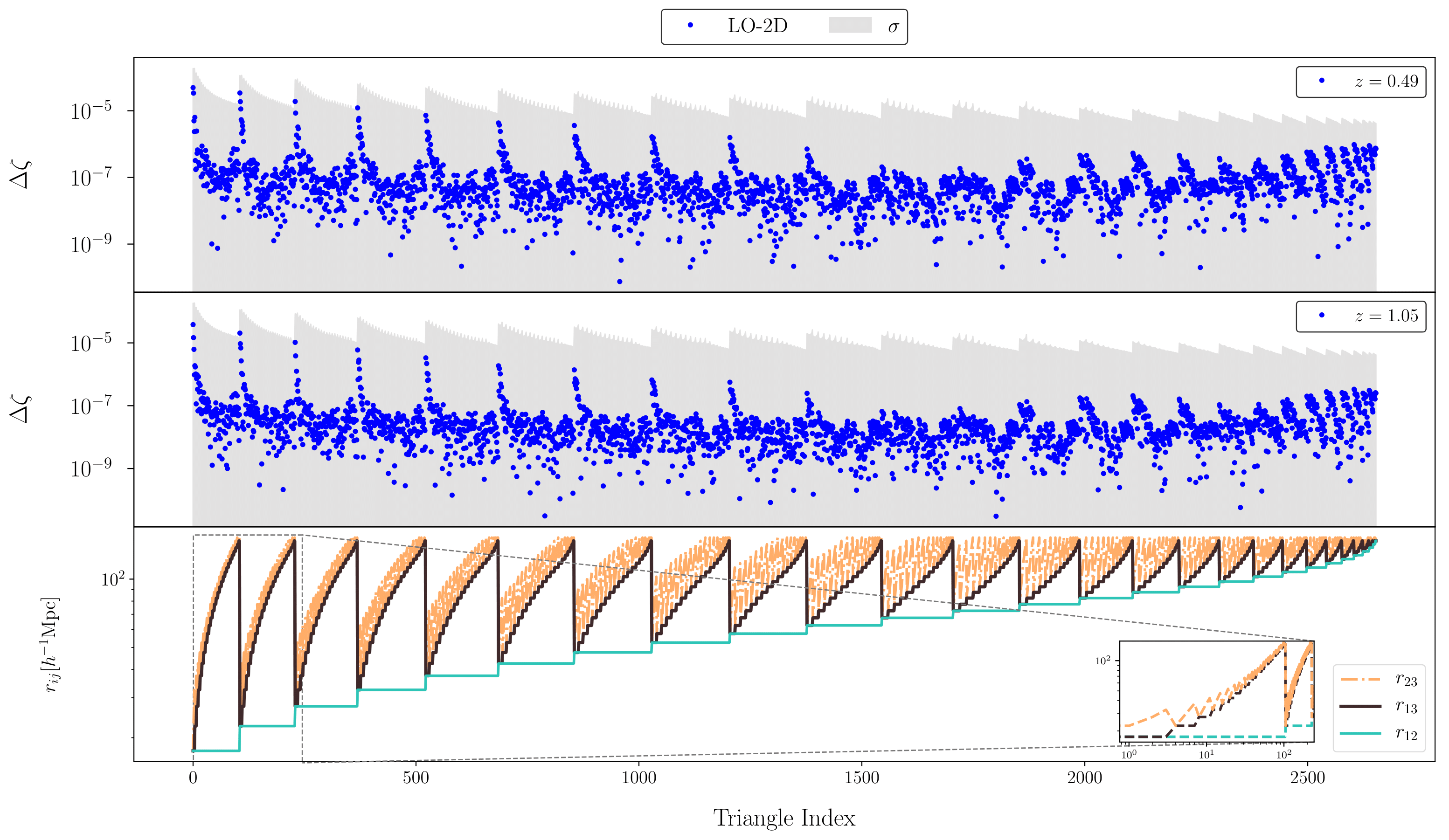}
\caption{Difference between LO-1D, \MG{considered as a benchmark}, and  LO-2D 3PCF models, $\Delta \zeta = \zeta^{\mathrm{LO-2D}} - \zeta^{\mathrm{LO-1D}}$ (blue curve), as a function of the triangle configurations identified by the side lengths. Top and central panels show the model predictions at $z= 1.05$ and at $z= 0.49$.
The grey band represents the 1-$\sigma$ Gaussian uncertainty. The bottom panel shows the different sides of the triangles as a function of the Triangle Index, \MG{using logarithmic y-axis}. \MG{The boxed plot represents, using logarithmic x- and y- axes, a zoomed-in area of the bottom panel including only separation distances $r_{12} \leq 17.5 h^{-1} \mathrm{Mpc}$, providing a closer look at the region of very small scales.}}
\label{fig:ftlog_meth_validation_LO}
\end{figure}

We now compare the LOs (computed via 1D- and 2D- FFTLog techniques) 3PCF models and NLO-SPT described in Sec. \ref{Method_Motivation_2DFFTLog}, using LO-1D as a reference case.
In Fig. \ref{fig:ftlog_meth_validation_LO}, we show the difference between the LO-2D and LO-1D 3PCFs, $\Delta \zeta$, as a function of the various triplets configuration, labelled and ordered as described in Sec. \ref{subsec:3PCF_models}. The top and bottom panels show the results obtained for $z=1.05$ and $z=0.49$, respectively. In both cases, the magnitude of the difference is small compared to the expected 1-$\sigma$ Gaussian error represented by the grey band. 
The largest discrepancies, the spikes in the blue curve, are found for the isosceles configurations $\eta_{\mathrm{min}} = 0$. \MG{These spikes are due to the numerical evaluation of the Legendre expansion of the bispectrum, that is the analogous of Eq. \ref{eq:leg_transf} in Fourier space. This expasion is performed by a Gauss-Legendre quadrature integration method with a fixed number of nodes. The amplitude of these peaks increases when decreasing the size of the triangles but never exceeds the Gaussian uncertainty (see Appendix \ref{AppendixB}).}
We conclude that our implementation, or the LO-2D model, which uses the 2D-FFTLog tool, fully agrees with the standard one, LO-1D, used as a benchmark. These results validate the 2D-FFTLog implementation and justify the adoption of this tool to generate a nonlinear 3PCF model.

\begin{figure}[H]
\includegraphics[width= 1. \textwidth]{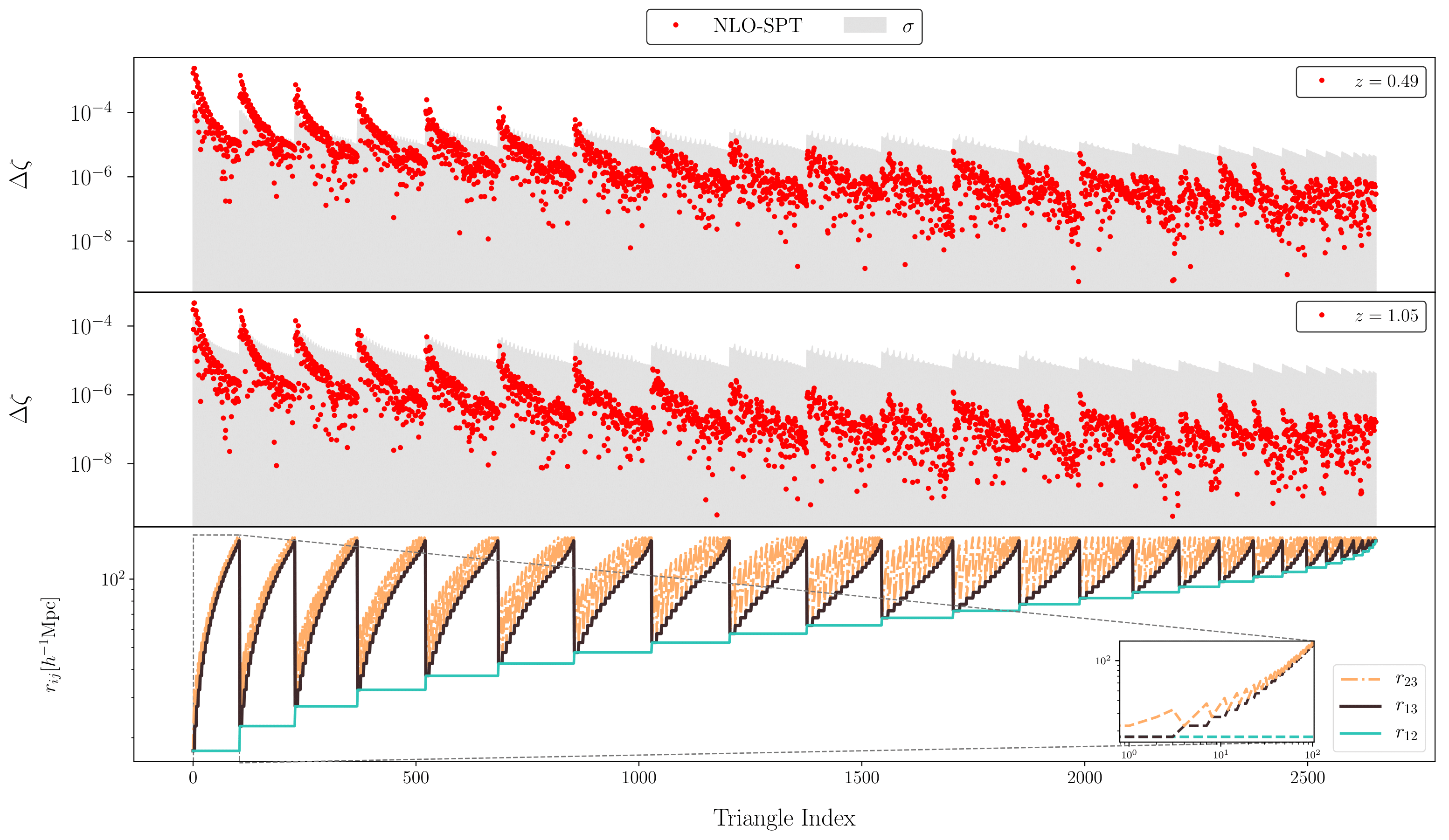}
\caption{Same as figure~\ref{fig:ftlog_meth_validation_LO} showing the difference between 3PCF models LO-1D and NLO-SPT (red curves).}

\label{fig:ftlog_meth_validation_NLO}
\end{figure}

Fig. \ref{fig:ftlog_meth_validation_NLO} compares the difference between the  
NLO-SPT and the LO-1D 3PCF models (red curve)  to the expected 1-$\sigma$ Gaussian uncertainty (grey band).
On large scales (i.e. to the right part of the plots), the differences are 
about ten times larger than in the LO-1D vs LO-2D case but still well below the expected Gaussian uncertainty, showing that nonlinear effects are small in this regime. Differences increase when moving to the left of the panels, i.e. on small scales, as expected. Differences between the models increase by up to two orders of magnitudes, indicating the importance of nonlinear contributions to the 3PCF signal. 

Moreover, the magnitude of the discrepancy increases at lower redshift for the same reason.
Superimposed on this trend, we still see peaks in correspondence of isosceles triangle configurations, confirming not only that the different techniques that reproduce the same model exhibit most differences on those configurations but also that differences between 3PCF models at different perturbative orders peak on small scales toward $\eta_{\mathrm{min}} = 0$.

We conclude that the NLO-SPT model significantly differs from the LO-1D on small scales. Which, however, does not guarantee that NLO-SPT is a better model. A point that we will address by comparing models with data in the next Sections.

The fact that the differences between models are larger than statistical errors for small triangle configurations indicates that model nonlinearities would dominate the error budget on small scales. And clearly illustrates the importance of going beyond the linear order 3PCF model.

\subsection{3PCF measurements}
\label{subsec:measure}

We performed 3PCF measurements on the DEMNUni simulation snapshots $z=0.49$ and $z=1.05$
using the estimators presented in Sec. \ref{sec:clustering_meas} over a wide range of scales: $r_{ij} = [17.5,132.5] \, h^{-1} \mathrm{Mpc} $ using the bin-width $\Delta r = 5 h^{-1} \mathrm{Mpc}$, \MG{thus using 27 radial bins}

\begin{figure}[H]
\includegraphics[width= 1 \textwidth]{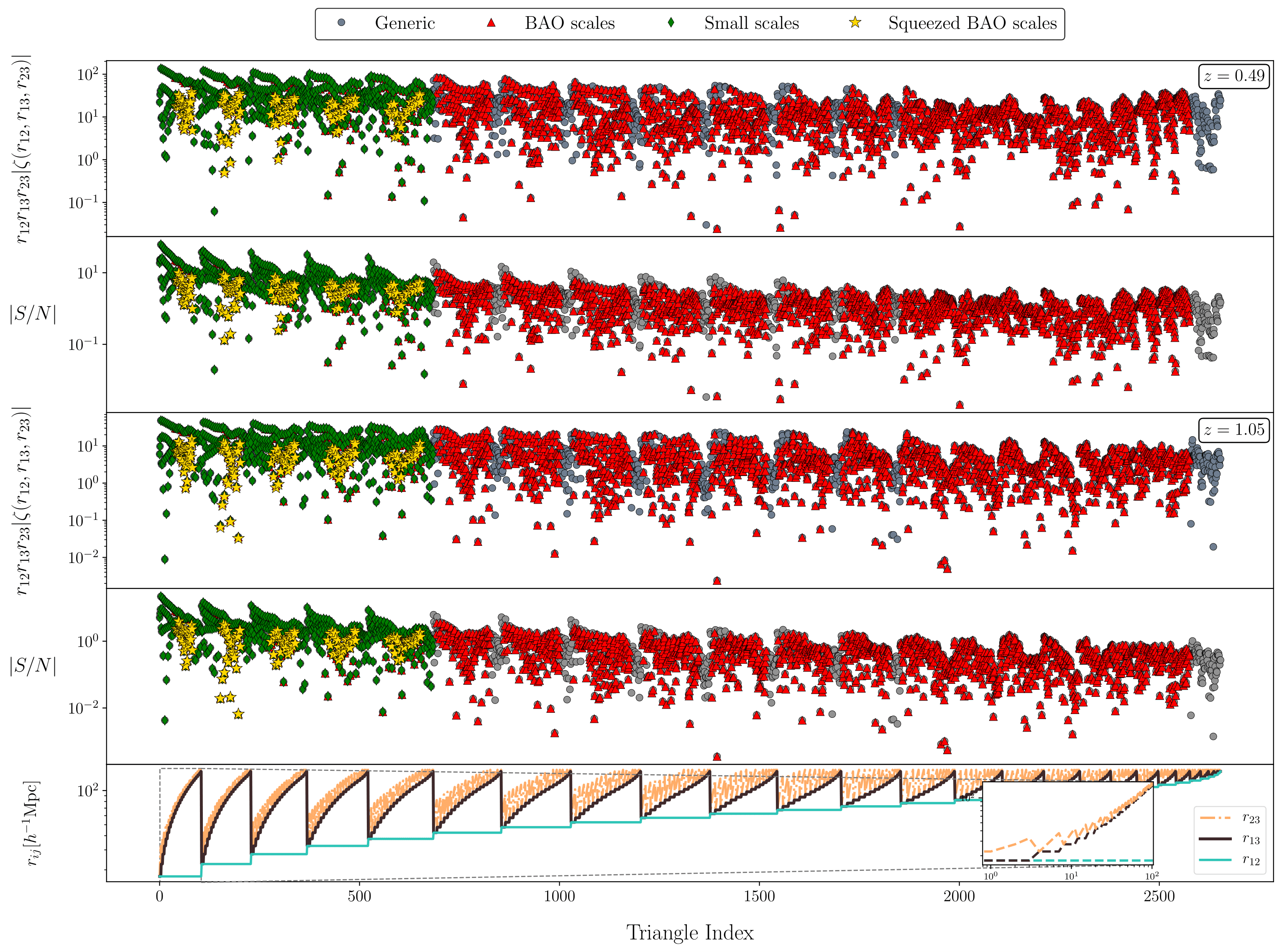}
\caption{Measurements of 3PCF from the DEMNUni snapshots at $z=0.49$
(top plots) and $z=1.05$ (bottom plots).
In each figure for different redshifts, the upper panel shows the 3PCF amplitude multiplied by the size of the sides of the triangles as a function of the triangle configuration, indicated on the X-axis. The bottom panel shows the signal-to-noise, assuming Gaussian errors.
Different colours and symbols are used for different triangle types, as indicated in the labels.
{In our classification scheme, as described in the text, it is possible for a triangle to have multiple classifications. For instance, a triangle classified as squeezed BAO would also fall into the category of small-scale triangles.}
First and second panels refer to $z= 0.49$, while third and fourth panels $z= 1.05$. Bottom plot shows the sides of the triangles as a function of the Triangle Index, ranging from $r_{\mathrm{min}} = 17.5 h^{-1} \mathrm{Mpc}$ and $r_{max} = 132.5 h^{-1} \mathrm{Mpc}$. The bin-width is $\Delta r = 5 h^{-1} \mathrm{Mpc}.$} 
\label{fig:measure}
\end{figure}

Every symbol in Fig. \ref{fig:measure} represents a 3PCF measurement
for triangles of all sizes, labelled in the X-axis, in the allowed range. Each panel is divided into two parts.
The upper plot shows the 3PCF amplitude multiplied by the triangle side lengths. The bottom plot shows the signal to noise, i.e. the 3PCF amplitude in units of the Gaussian error: $\zeta/\sigma$. Different symbols (and colours) are used for different triplet types. Green dots indicate small-scale triplets, defined as configurations for which $r_{12} \leq 40 h^{-1} \mathrm{Mpc}$. Red triangles identify BAO triplets that encompass the BAO scale, the latter defined as the scale in which BAOs provide their typical wiggling features to the correlation functions. Practically, for BAO configurations, we consider triangle configurations with at least one side in the range $[17.5 - 117.5 h^{-1} \mathrm{Mpc}]$. Yellow stars indicate the squeezed BAO configurations, i.e. triplets with one side much smaller than the other two, whose at least one is in the BAO regime. The grey dot symbol is used for all other cases. The signal to noise is typically above unity, quite insensitive to the triangle size except on small scales where it peaks.

\subsection{3PCF models vs. data: overview}
\label{subsec:modelcomparison}

\begin{figure}[H]
\includegraphics[width= 0.5 \textwidth]{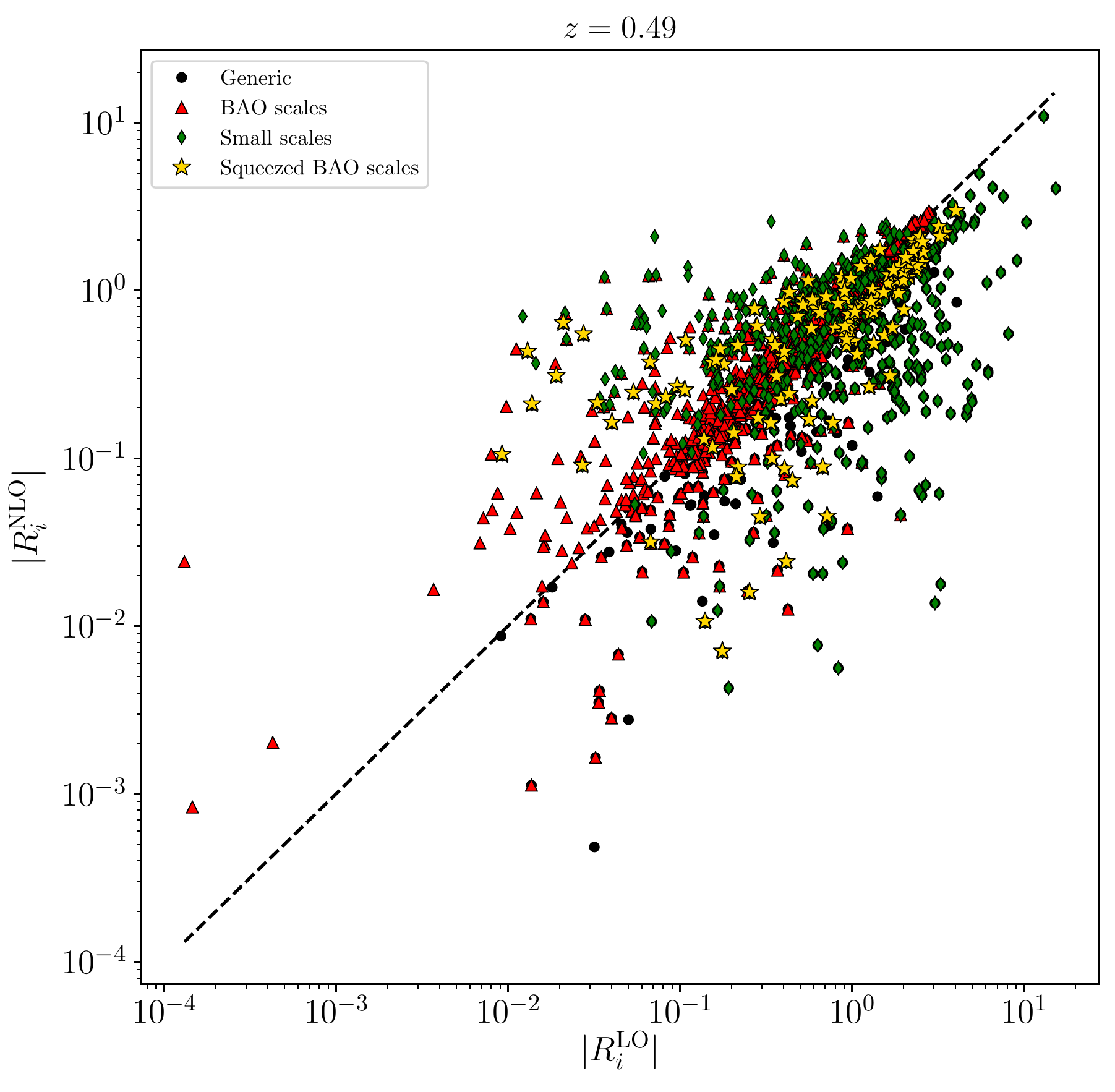}
\includegraphics[width= 0.5 \textwidth]{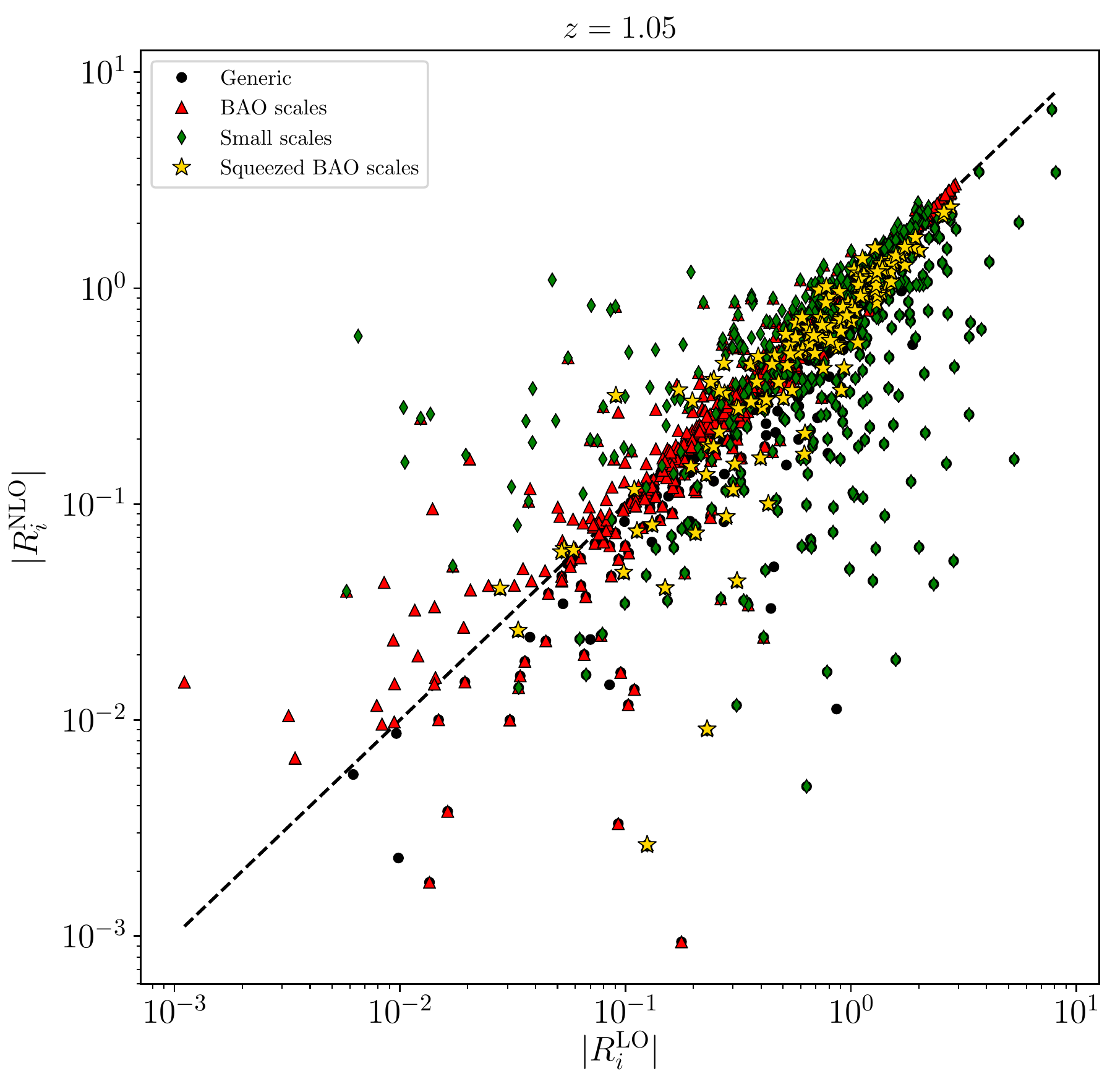}
\caption{Comparison between normalised absolute values of residuals of SPT-NLO and LO-2D models. The LO-2D residuals are displaced in the X-axis, while the SPT-NLO case is displaced in the Y-axis. The left plot refers to $z= 0.49$, while the right plot refers to $z= 1.05$. Different symbols focus on BAO scales (red triangles), small scales (green rhombuses) and \MG{squeezed} BAO scales (yellow stars). }
\label{fig:residuals}
\end{figure}

We compare the NLO-SPT and the LO-2D  3PCF models to the same set of measurements performed on the DEMNUni datasets. The scatterplot in Fig.~\ref{fig:residuals} compares the absolute values of normalised residuals in Eq.~\eqref{eq:res} of both models for each triangle configuration for the NLO-SPT  (Y-axis) and LO-2D (X-axis) models in the two snapshots. We use the same symbols and colors as in Fig.~\ref{fig:measure} to identify different types of triangle configurations. In Fig.~\ref{fig:residuals}, we show the normalised absolute value of residuals of NLO-SPT on the Y-axis and LO-2D on the X-axis for both redshift cases. In the left and right panels, respectively $z=0.49$ and $z=1.05$, different symbols and colours are used to represent the different configurations shown in Fig.~\ref{fig:measure}. Deviations under the diagonal indicate the SPT-NLO model behaves better with respect to the LO-2D model. Deviations above the diagonal indicate the opposite case.


We stress that a clear and expected feature is the redshift dependence: the matching between model and data improves with the redshift for both the NLO-SPT and the LO-2D cases. At $z=1.05$ only for a handful of small triangle configurations, the difference between the model and the measured 3PCF exceeds 3-$\sigma$ significance. And these are mostly LO-2D predictions. At $z=0.49$, the number of 3-$\sigma$ outliers increases significantly, and, more generally, the amplitude of the data-model mismatch systematically increases on all scales. We also represented different sub-sets of configurations with different colours and symbols: small scales, BAO configurations, and squeezed BAO scales.
To quantify the relative agreement between the LO-2D and SPT-NLO models, we report the average difference between the absolute value of normalised residuals (see Eq. \ref{eq:delta_res}) - for each considered sub-set - in Tab. \ref{tab:summary_residuals_confs}. Negative values of the average difference of the absolute values of normalised residuals indicate the NLO-SPT's predictions are closer to the simulated dataset with respect to the LO-2D's case; positive values indicate the opposite. The whole set of measured triangle configurations shows negative values of $\left< \Delta R \right>$. Small scales indicate an even more negative value of that quantity, showing the NLO-SPT model exhibits a smaller difference with the data compared to the LO-2D case. Similarly, for the BAO scales, for which the cited difference assumes a slightly negative value despite being consistent with zero. As a preliminary overview, these cases indicate the contribution of the SPT-NLO model better matches the simulated dataset. Considering the squeezed BAO scales, $\left< \Delta R \right>$ is negative, and it represents the most significant deviation from the zero value considering the error, meaning the SPT-NLO model improves the match with the dataset concerning the LO-2D model. At this stage, we did not consider the correlation between measurements, ignoring the effect of the non-diagonal terms in the covariance matrix. We will properly quantify the performance of models in Sec \ref{subsec:fitmodels}. We will focus on the particular squeezed isosceles configurations in the next Sec.

\begin{table}[H]
    \begin{center}
    \renewcommand{\arraystretch}{1.4}
    \vspace*{1 cm}
    \begin{tabular}{c|c|c}
         \hline
         Configurations & $z=0.49$ & $z=1.05$ \\ 
         \hline
         \ All configurations & $\left< \Delta R \right> = -0.129 \pm 0.038$ & $\left< \Delta R \right> = -0.059 \pm 0.039$ \\
         \ BAO scales & $\left< \Delta R \right> = -0.009 \pm 0.045$ & $\left< \Delta R \right> = -0.006 \pm 0.045$ \\
         \ Small scales & $\left< \Delta R \right> = -0.431 \pm 0.076$ & $\left< \Delta R \right> = -0.188 \pm 0.076$ \\
         \ Squeezed BAO & $\left< \Delta R \right> = -0.262 \pm 0.163$ & $\left< \Delta R \right> = -0.132 \pm 0.163$ \\
         \hline
    \end{tabular}
    \end{center}
    \caption{Average difference between the absolute values of normalised residuals of SPT-NLO and LO-2D models. Redshifts are listed in columns two to three, while all configurations, BAO, small scales and squeezed BAO scales are listed in rows two to five. }

    \label{tab:summary_residuals_confs}
\end{table}

\subsection{3PCF models vs. data: the BAO scale}
\label{subsec:BAO}

A key feature in clustering statistics is represented by the imprint of the baryons, the large-scale structures that can be used to trace the expansion history of the Universe with per cent precision. BAOs in two-point statistics have become a standard and very effective cosmological probe. However, the BAO feature can be, and has been, detected in the galaxy 3-point correlation function, too \citep{Moresco2021, Slepian:2016kfz}. As for the 2-point statistics, its use for precision cosmology depends on the ability to model nonlinear effects, i.e. to go beyond linear order-based models. Indeed, correct modelling of the BAO is important in full shape fits, as throughout, we are able to get cosmological information.

In this Sec., we, therefore, focus on the BAO scale, and more precisely on squeezed isosceles triangle configurations encompassing the BAO scale, to assess the performance of the  NLO-SPT \MG{and NLO-EFT } models in comparison to LO's predictions.


\begin{figure}[ht!]
    \includegraphics[width=.5\textwidth]{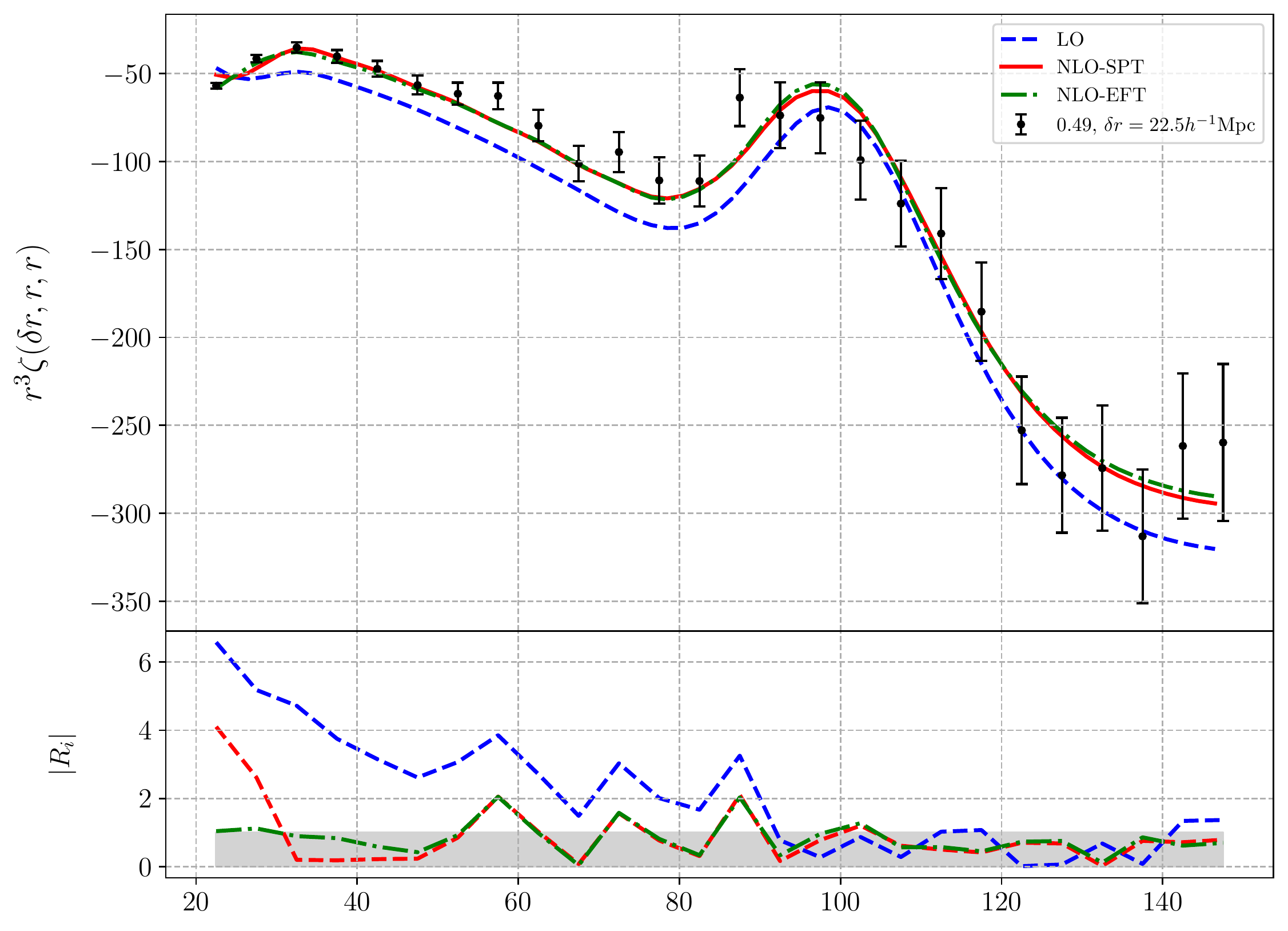}
    \includegraphics[width=.5\textwidth]{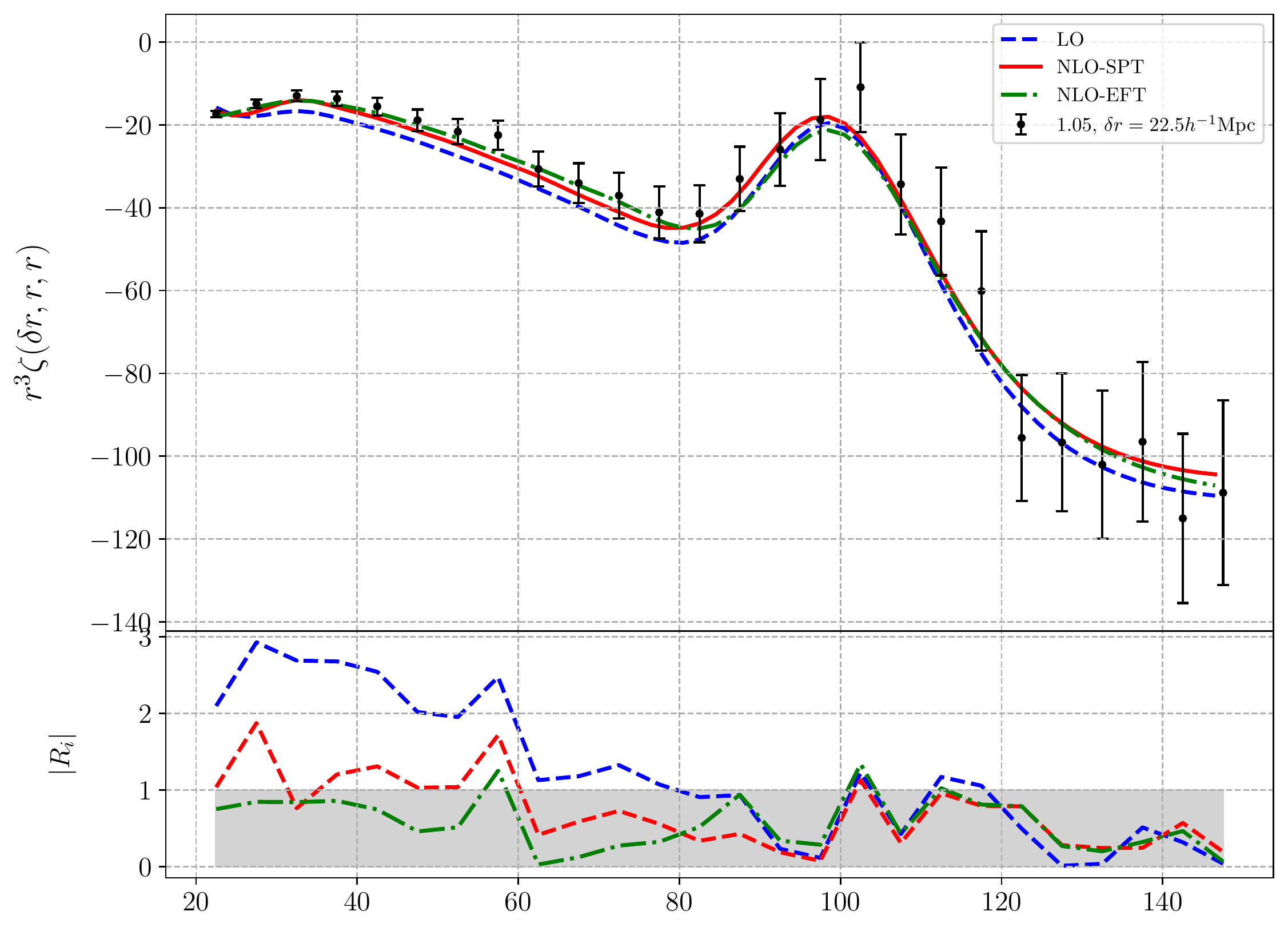}
    \caption{\MG{\textit{Top:}} 3PCF as a function of the two side lengths $r$ of a set of isosceles triangles, with the third side of length \MG{$\delta r = 22.5 h^{-1} \mathrm{Mpc}$}. The 3PCF measurements are shown as black dots, and the error bars indicate the 1 $\sigma$ Gaussian uncertainty. The continuous red curve and the dashed blue ones show the predictions of the NLO-SPT of the LO-2D models, respectively. Left and right panels illustrate the results at 
    $z= 0.49$ and at $z= 1.05$. \MG{\textit{Bottom:} normalised residuals for $z=0.49$ (left), $z=1.05$ (right), as defined in Eq. \ref{eq:res}, for the three models depicted by the same colours and shapes of the top panel.}}
    \label{fig:squeezed_BAO_z1}
\end{figure}

\begin{table}[H]
    \begin{center}
    \renewcommand{\arraystretch}{1.4}
    \vspace*{1 cm}
    \begin{tabular}{c|c|c|c|c}
         \hline
         Redshift & $c_0 \ \big [ h^{-2} \mathrm{Mpc^2} \big ]$ & $c_1 \ \big [ h^{-2} \mathrm{Mpc^2} \big ]$ & $c_2 \ \big [h^{-2} \mathrm{Mpc^2} \big] $ & $c_3 \ \big [ h^{-2} \mathrm{Mpc^2} \big] $ \\ 
         \hline
         $z = 0.49$ & $2.25 \pm 1.68$ & $-4.20 \pm 1.73$ & $-1.91 \pm 2.37$ & $1.65 \pm 3.25$ \\
         $z = 1.05$ & $3.68 \pm 1.90$ & $-4.81 \pm 1.96$ & $-7.38 \pm 2.68$ & $11.45 \pm 3.65$ \\
    \end{tabular}
    \end{center}
    \caption{\MG{Best fit values for the four parameters of the NLO-EFT model for the squeezed BAO configurations, as depicted in Fig. \ref{fig:squeezed_BAO_z1}.}}

    \label{tab:EFT_BAO}
\end{table}

In Fig.\ref{fig:squeezed_BAO_z1}, we compare the measured and the predicted 3PCFs for isosceles triangles with one side length fixed to \MG{$\delta r = 22.5$ $h^{-1} \mathrm{Mpc}$}, well within the nonlinear regime, and the other two equal side of increasing length $r$, shown on the X-axis. 3PCF estimated are represented by the black dots with their 1 $\sigma$-Gaussian error bars. Models' predictions are shown by the two curves: dashed blue for the LO-2D case, continuous red for the NLO-SPT \MG{and dashed-dot green for the NLO-EFT}. 
At low redshift (left panel), the NLO-SPT \MG{and NLO-EFT} models outperform LO-2D on all scales. In particular, they reproduce the shape of the BAO peak much better, indicating that the nonlinear effects responsible for the widening of the peak are correctly accounted for.
At larger redshifts, the \MG{three} models behave almost identically --- \MG{particularly on larger sides} --- which does not come as a surprise. \MG{In regard to the comparison between the NLO-EFT model and the NLO-SPT model, incorporating four parameters leads to a significant improvement in data matching, particularly at small scales, as shown by the normalised residual depicted in the bottom panel of Fig. \ref{fig:squeezed_BAO_z1}.}
We notice that at this redshift and on these scales, the 3PCF signal is very small (notice the difference in the Y-axis scale) and highly correlated, which explains why, as shown in the next Sec.,  
the statistical significance of the mismatch as quantified by the 
$\chi^2$ analysis is rather small.




\subsection{3PCF models vs. data: quantitative analysis}
\label{subsec:fitmodels}

We now present the results of the $\chi^2$ analysis introduced in Section \ref{sec:matching}.
In addition to LO-2D and NLO-SPT, we also consider the nonlinear NLO-EFT described in Section \ref{subsec:EFT}. In these analyses, we have considered all triangles with side length $r_{12}$ in the range $[17.5, \MG{132.5}] \, h^{-1} \mathrm{Mpc}$, splitting our analysis in two cases: $\eta_{\mathrm{min}} = 2$ and $\eta_{\mathrm{min}} = 0$.

\begin{figure}[!ht]
\begin{center}
    
    \includegraphics[width=1.\textwidth]{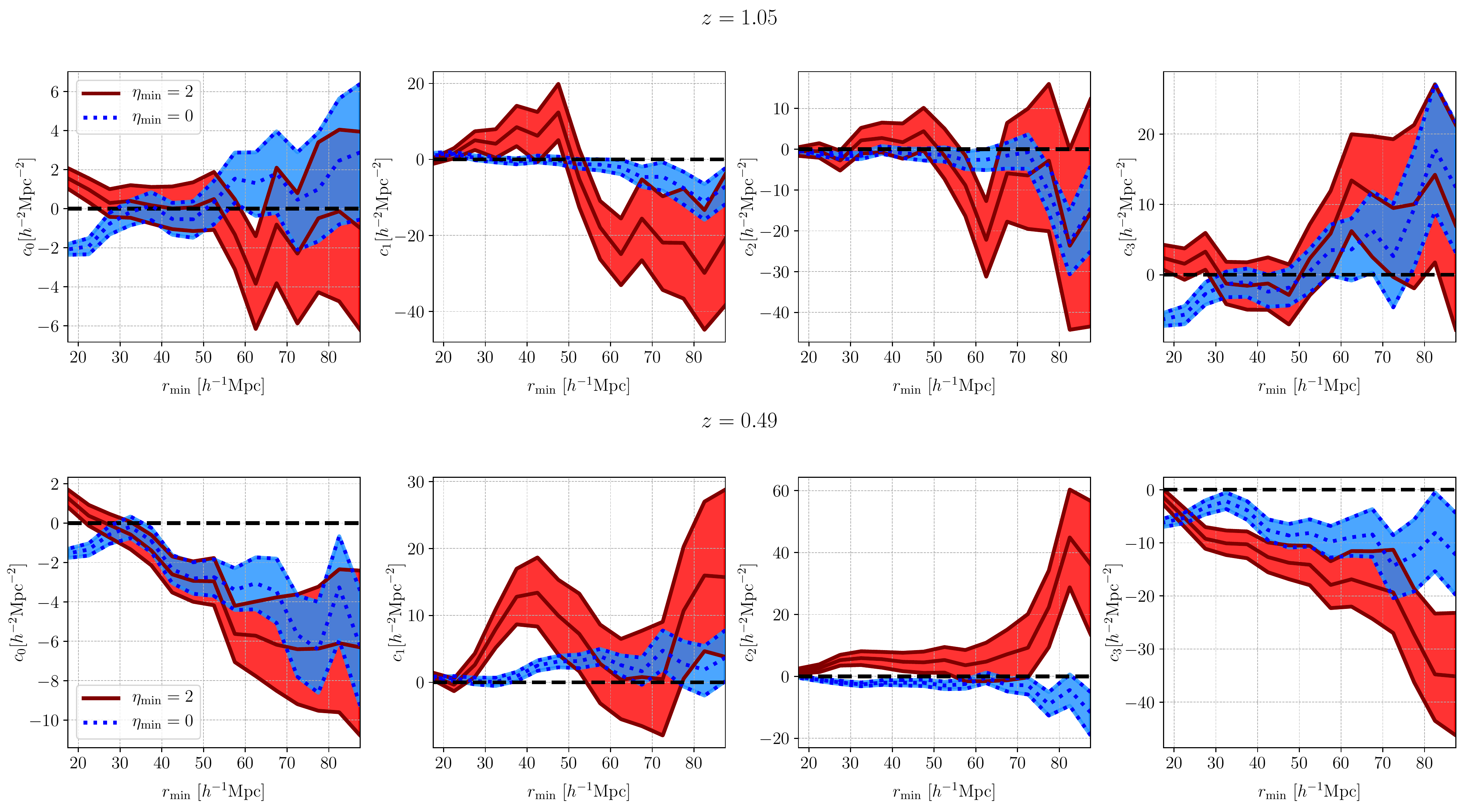}

    \caption{\MG{Best fit values for the four NLO-EFT free parameters (left to right) from the reduced $\chi^2$ analysis, as a function of 
     $r_{\mathrm{min}}$. The red continuous band refers to $\eta_{\mathrm{min}} = 2$, the blue dotted band to $\eta_{\mathrm{min}} = 0$. Top and bottom panels show the results for the $z  = 0.49$ and $z=1.05$ snapshots, respectively.}}
    \label{fig:EFT_fit}
\end{center}
\end{figure}
\begin{figure}[H]
    \includegraphics[width=1.\textwidth]{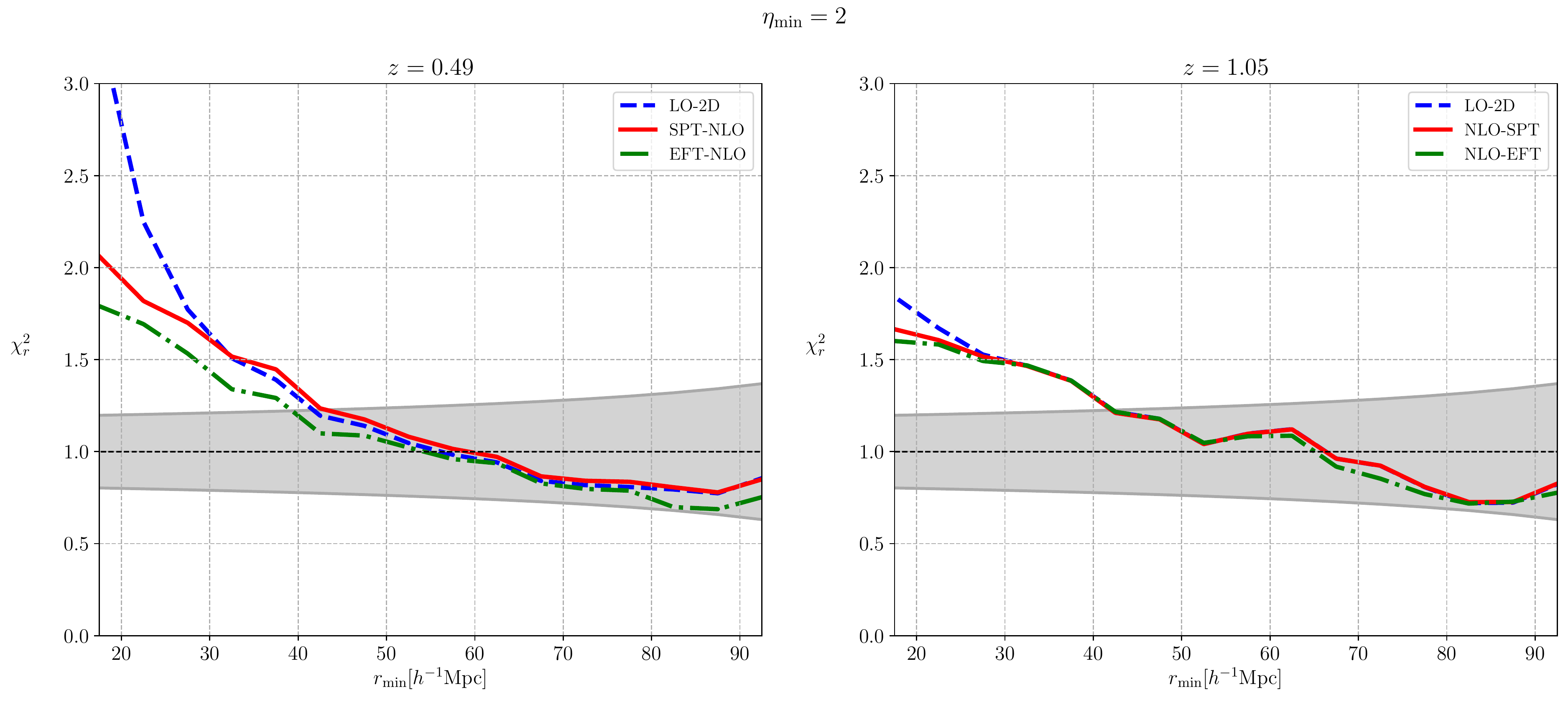}
    \caption{\MG{Cumulative reduced chi-square $\chi^2_r$ in the range 
    $[r_{\mathrm{min}},132.5] h^{-1} \mathrm{Mpc}$ where $r_{\mathrm{min}}$ spanning in the range $[17.5, 92.5] h^{-1} \mathrm{Mpc}$, as a function of 
    $r_\mathrm{{min}}$ for the case $\eta_{\mathrm{min}} = 2$. 
    continuous red and dashed blue curves refer, respectively, to the NLO-SPT and LO-2D models. The grey band represents a 99.7\% confidence level assuming Gaussian statistics. The left and the right panels refer, respectively, to redshift $z=0.49$ and $z= 1.05$}}
    \label{fig:chi_eta_2}
\end{figure}

\textcolor{black}{The reason why we ran the analysis with different values of $\eta$ is motivated by the measurement technique adopted. Indeed we measure the 3PCF multipoles first, up to $\ell_{max} = 10$, and then combine them. This is not a problem in principle as long as we repeat the same operation consistently in the model. The complication arises for those combination $\lbrace r_{12},r_{13} \rbrace$ that allows the third side $r_{23}$ to span from very large to very small, highly nonlinear scales. For such cases, corresponding to the choice of $\eta=0,1$, the multipole series is slowly convergent due to the steep shape of the 3PCF as $r_{23} \rightarrow 0$. All the resummed 3PCF coming from these multipoles are then dominated by the contribution of these squeezed configurations, which are extremely difficult to properly model. In principle, a standard estimator of the 3PCF would allow to filter these contributions more efficiently.}
In order to compare NLO-EFT predictions with the ones from other models, we inferred the four model parameters for both cases of choice of $\eta_{\mathrm{min}}$ and redshifts as a fixing the maximum scale to \MG{$r_{\mathrm{max}} = 132.5$} and varying the minimum scale of the fit $r_{\mathrm{min}}$ in the interval $[17.5, \MG{87.5}] \, h^{-1} \mathrm{Mpc}$, i.e. the range from mild to strong nonlinear scales in which EFT contributions are supposed to be relevant. The results are shown in Fig. \ref{fig:EFT_fit}. 
The $\eta_{\mathrm{min}} = 0$ case, due to the larger number of triangle configurations, exhibits smaller posterior contours with respect to the $\eta_{\mathrm{min}} = 2$ case. This plot is to be interpreted cumulatively from right to left: on large scales where the nonlinear effects are small, their values are consistent with zero within the error bars. Moving towards smaller scales and considering an increasingly large number of progressively smaller triangles, the parameters of the best-fitting models significantly depart from zero at both redshifts and $\eta_{\mathrm{min}}$ choice. The estimation of $c_2$ benefits from the larger number of configurations in the case $\eta_{\mathrm{min}} = 0$. Using the inferred EFT parameters at each $r_{\mathrm{min}}$ for the NLO-EFT model, we present the performance of the goodness-of-fit of all the models listed in Tab. \ref{tab:models_resume}. The results are shown in Fig. \ref{fig:chi_eta_2} in the form of cumulative $\chi^2$: from right to left, each point of the curves indicates the reduced $\chi^2$ value obtained by including all triangles with 
$\eta_{\mathrm{min}} = 2$ and $r_{12}$ in the range $[r_{\mathrm{min}}, \MG{132.5}] \, h^{-1} \mathrm{Mpc}$, where the $r_{\mathrm{min}}$ value is to be read on the X-axis.

The leftmost point of the curves indicates the $\chi^2$ value obtained when considering all triangle configurations with  $\eta \geq 2$.
The cumulative  $\chi^2_r$ value of each model can be compared with the corresponding  99.7\% confidence interval for Gaussian statistics, represented by the grey band. The reduced $\chi^2$ is well within the grey strip for most of the scales above $40 h^{-1} \mathrm{Mpc}$, indicating those model uncertainties provide a minor contribution to the 3PCF total error budget.  For this purpose, we normalised the covariance matrix in Eq. \ref{eq:cov_2} to have a meaningful goodness-of-git around unity on large scales in which models are expected to work well. 
This is not the case for scales below $40  h^{-1} \mathrm{Mpc}$, where nonlinear effects are large and cannot be fully captured by perturbative models. We observe that for scales below $30  h^{-1} \mathrm{Mpc}$ at both redshifts, the difference between the NLOs models and the LO-2D model is appreciable, particularly for $z= 0.49$, but limited to scales in which uncertainties are known to be underestimated, and goodness-of-fit is far from being close to unity. These are also the scales, however, on which  the NLO-SPT and NLO-EFT provide a better fit to the data than the LO-2D one. The small difference between the models is slightly deceptive, and partly due to the fact that having considered cumulative statistics, the $\chi^2_r$ is significantly contributed by those large triangles in which the two models provide similar (good) predictions. The fact that LO and NLO models perform similarly in the $\eta_{\mathrm{min}} = 2$ case reflects the fact that these are triangle configurations for which the number of scales in the small-scale regime is limited by the condition.

Therefore we decided to explore the more challenging case of $\eta_{\mathrm{min}} = 0$ for which the number of triangle sides in the small scale range is as large as possible, ranging in $[r_{\mathrm{min}}, \MG{132.5 h^{-1} \mathrm{Mpc}}]$.

\begin{figure}[H]
    \includegraphics[width=1.\textwidth]{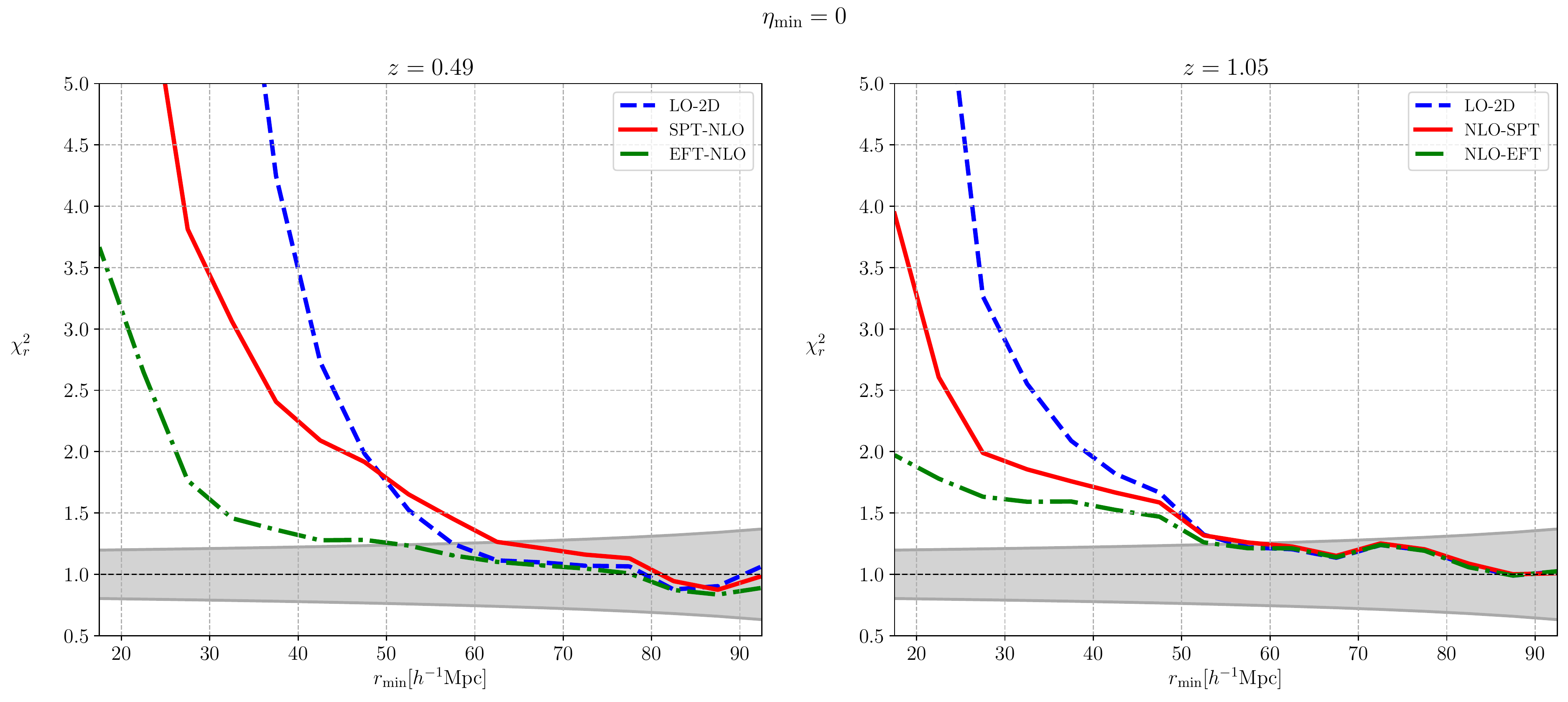}
    \caption{\MG{Same as Fig. \ref{fig:chi_eta_2} fore the case $\eta_{\mathrm{min}} = 0$. Model predictions for NLO-EFT, NLO-SPT and LO-2D are shown with
    green dot-dashed, reds continuous and blue dashed curves, respectively.}}
    \label{fig:chi_eta_0}
\end{figure}

Fig. \ref{fig:chi_eta_0}, in which cumulative  $\chi^2_r$ of all models are represented, shows that while the goodness of the fit for  both next-to-leading models, NLO-SPT and  NLO-EFT, have significantly worsened with respect to the $\eta_{\mathrm{min}} = 2$ configuration case.
And yet they both outperform the LO-2D model on small scales at both redshifts. Here, the improvement coming from the SPT-NLO is more significant and, as in the previous case at $\eta_{\mathrm{min}} = 2$, bounded at scales below \MG{$60 h^{-1} \mathrm{Mpc}$}, and where goodness-of-fit is far from being close to the expected value. This is influenced by the underestimated estimation of uncertainties in this regime. The comparison between these different models informs us that the use of nonlinear models is mandatory to analyse the 3PCF signal from  $\eta_{\mathrm{min}} < 2$ triangle configurations. And that even on small scales, where the quality of the fit degrades, the systematic errors they introduce are significantly smaller than bias of the LO models.

Although the goodness of the NLO-EFT fit is bound to be superior to  the  NLO-SPT one, the corresponding reduced $\chi^2$ values are quite similar in the two cases.



\begin{figure}[H]
    \includegraphics[width=1.\textwidth]{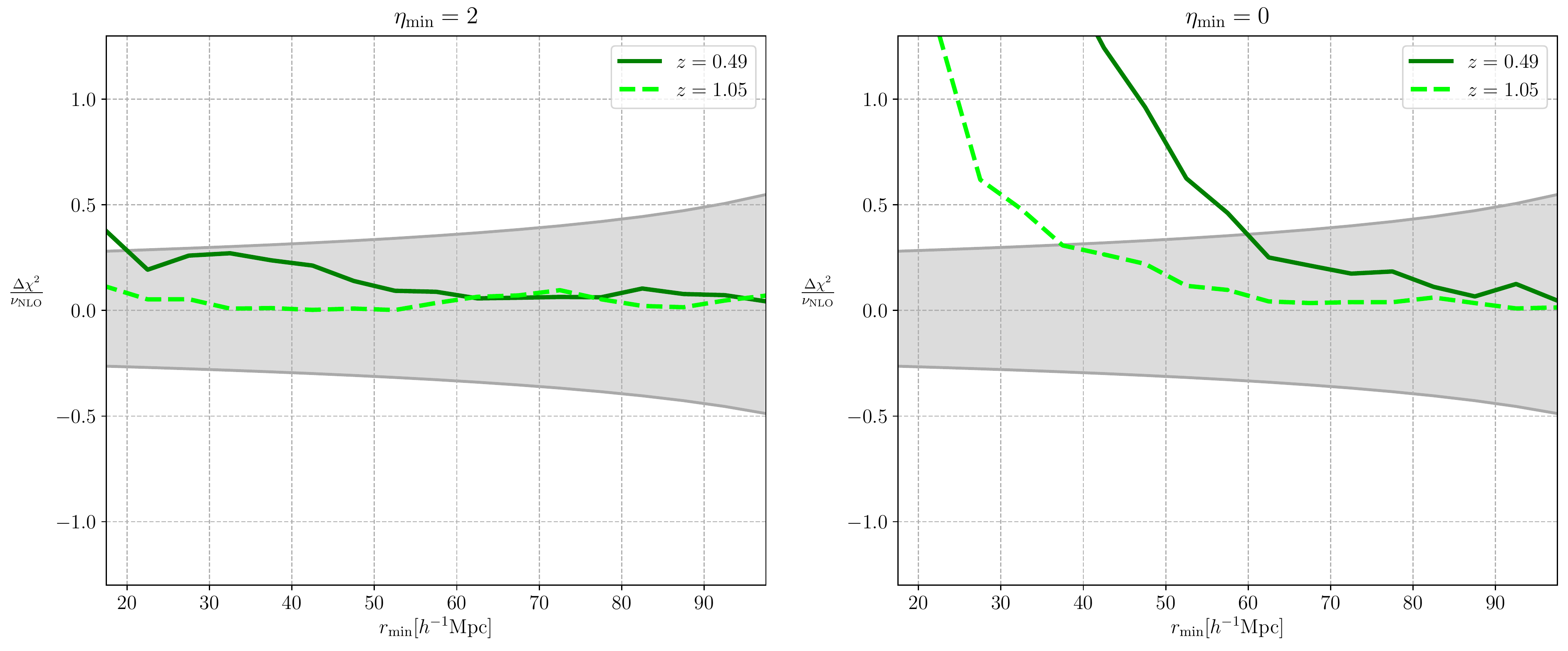}
    \caption{\MG{Cumulative chi-square difference between NLO-SPT and NLO-EFT models divided by the NLO-SPT degree of freedom. Dotted lime line refers to $z= 1.05$, the solid green one refers to $z= 0.49$. From left to right, respectively, the case  $\eta_{\mathrm{min}} = 2$ and  $\eta_{\mathrm{min}} = 0$. The grey band represents a $99.7\%$ confidence level coming from the propagation of errors on Eq. \ref{eq:diff_chi}.}}
    \label{fig:delta_chi_EFT}
\end{figure}

In order to quantify the improvement coming from the adoption of the NLO-EFT model with respect to SPT-NLO, we studied the cumulative chi-square difference normalised by NLO-SPT degrees of freedom as presented in Eq. \ref{eq:diff_chi}. Results are shown in Fig. \ref{fig:delta_chi_EFT}. 
Differences above the grey band mean the improvement coming from NLO-EFT is significant. None of the differences is below the grey band, meaning the model benefits or - at least - is not worsened by the adoption of four extra fitting parameters, as expected. The case $\eta_{\mathrm{min}} = 2$  shows the NLO-EFT models  \MG{show an almost insignificant improvement over the NLO-SPT model}, at both redshifts. \MG{On the other hand}, the case $\eta_{\mathrm{min}} = 0$ shows the improvement coming from NLO-EFT model is significant in the \MG{\sout{very}} small-scale regime, confined to scales below \MG{$60, 40 h^{-1} \mathrm{Mpc}$, respectively at $z = 0.49, 1.05$}. However, the improvement holds in the regime in which Gaussian uncertainties are supposed to be underestimated \citep{Veropalumbo2022} so that the goodness of fit is far from being close to the expected value in this regime. What is surprising is the contribution from the adoption of EFT counterterms in a configuration space analysis as we focused on this work, and in a Fourier space analysis as done in \citep{Alkhanishvili:2021pvy}. Bispectrum analyses at next-to-leading order have shown that EFT models are crucial to model the matter signal up to $k = 0.16 - 0.19 \mathrm{Mpc}^{-1} h$ depending on the binning of the data, being a significant improvement with respect to the SPT model (see Fig. 7 of \citep{Alkhanishvili:2021pvy}). In our analysis, the improvement of the EFT-NLO model compared to SPT-NLO is much less significant, although the quantification of the improvement is affected by the estimation of the uncertainties that differ in both analyses. This seems to reflect that departures from the SPT model parametrised by adding four extra EFT terms in Fourier space are converted into very small-scale contributions in configuration space, for which both models fail to give a clear representation of the data.

To summarize our results we list, in Tab. \ref{tab:summary_fit_eta0} and \ref{tab:summary_fit_eta2} , the signal-to-noise values 
obtained by comparing the model 3PCF to the data (column 5) at 1) $r_{\mathrm{min}} = 40 h^{-1} \mathrm{Mpc}$ for $\eta_{\mathrm{min}} = 0$ and 2) $r_{\mathrm{min}} = 20 h^{-1} \mathrm{Mpc}$ for $\eta_{\mathrm{min}} = 2$. For both choices, the NLOs models exhibit smaller goodness-of-fit, meaning in the small scale range, the adoption of NLOs models is required to model the dataset properly. Comparing both choices, smaller goodness-of-fit corresponds to smaller $S/N$ values, even if choice 2) considers smaller minimum scales than 1). We stress the agreement between models and data, for all the models considered in this work in the nonlinear regime, is affected by the estimation of uncertainties \citep{Veropalumbo2022}.

\begin{table}[ht!]
    \begin{center}
    \renewcommand{\arraystretch}{1.4}
    \begin{tabular}{c|c|c|c}
         \hline
         Model & Redsfhit & $\chi_\mathrm{r}^2$ & $S/N$ \\
         \hline
         LO-2D  & 0.49 & 2.73 & 19.58 \\  
         \ EFT/NLO-SPT  & 0.49 & 2.09 & 19.58  \\   
         \ LO-2D  & 1.05 & 1.82 & 13.07 \\  
         \ EFT/NLO-SPT  & 1.05 & 1.66 & 13.07 \\ 
         \hline
    \end{tabular}
    \end{center}
    \caption{Signal-to-noise $[S/N]$ at $r_{\mathrm{min}} = 40 h^{-1} \mathrm{Mpc}$ and $\eta_{\mathrm{min}} = 0$ and the $\chi^2$ comparison between the model 3PCFs considered in this work, and listed in column 1, and the measurements performed on two snapshots of the DEMNUni simulations at the redshifts listed in column 2.}
    \label{tab:summary_fit_eta0}
\end{table}

\begin{table}[ht!]
    \begin{center}
    \renewcommand{\arraystretch}{1.4}
    \begin{tabular}{c|c|c|c}
         \hline
         Model & Redshift & $\chi_\mathrm{r}^2$ & $S/N$ \\
         \hline
         \ LO-2D  & 0.49 & 2.24 & 9.55 \\  
         \ EFT/NLO-SPT  & 0.49 & 2.06 & 9.55  \\   
         \ LO-2D  & 1.05 & 1.67 & 12.05 \\  
         \ EFT/NLO-SPT  & 1.05 & 1.61 & 12.05 \\ 
         \hline
    \end{tabular}
    \end{center}
    \caption{Same as Tab. \ref{tab:summary_fit_eta0}, but  $r_{\mathrm{min}} = 20 h^{-1} \mathrm{Mpc}$ and $\eta_{\mathrm{min}} = 2$}

    \label{tab:summary_fit_eta2}
\end{table}

Focusing on the BAO scales, the differences between the leading and next-to-leading order models are also significant. The comparison, described in the previous Section, considers a wide range of scales $r = [22.5,132.5] h^{-1} \mathrm{Mpc}$ and it is summarised in Tab \ref{tab:summary_fit_BAO}. Especially at $z= 0.49$, for the configuration we used as a test, the chi-square value is high, but, along with the case at $z=1.05$, it is shown that the SPT-NLO model provides a significant improvement on squeezed BAO configuration. We stress the significance of the mismatch between models and data depends on the adopted model, and it is affected by having adopted a Gaussian model for the errors and their covariance.


\begin{table}[H]
    \begin{center}
    \renewcommand{\arraystretch}{1.4}
    \begin{tabular}{c|c|c}
         \hline
         Model & Redshift &  $\chi^2_r$ \\
         \hline
         \ LO-2D & 0.49 &  27.6  \\  
         \ NLO-SPT & 0.49 & 11.9 \\ 
         \ \MG{NLO-EFT} & 0.49 & 2.9 \\ 
         \ LO-2D & 1.05 & 4.4  \\  
         \ NLO-SPT & 1.05 & 2.3  \\ 
         \ \MG{NLO-EFT} & 1.05 & 1.0 \\ 
         \hline
    \end{tabular}
    \end{center}
    \caption{Reduced $\chi^2$ values obtained when comparing measured and model 3PCF for triangle configurations that encompass the BAO scale (column 3).
    The models are listed in column 1, and the redshift of the simulation snapshot is shown in column 2. The analysis is described in Sec. \ref{subsec:BAO}.}
    \label{tab:summary_fit_BAO}
\end{table}


\section{Discussion and conclusion}

\label{sec:summary}

This work represents a first step in modelling 
3-point statistics in configuration space with a precision comparable to that of the nonlinear galaxy bispectrum model in Fourier space, focusing on the matter 3-point correlation function,
We have proposed two new next-to-leading perturbation theory models for the matter 3-point correlation function
(dubbed NLO-SPT and NLO-EFT in Table \ref{tab:models_resume}), compared their improvement over the leading order model (LO-2D) and gauged their performance against the matter 3PCF measured in two snapshots of the 
DEMNUni simulation at redshifts $z=0.49$ and $z=1.05$.

\begin{itemize}
    \item All the 3PCF models presented in this work are based on the perturbation theory bispectrum model \citep{Scoccimarro98} and rely on the 2D-FFTLog to obtain the multipoles of the 3-point function. The procedure is computationally intensive since functions have to be evaluated on large, $256 \times 256 \times 51$ grids to reach the required accuracy that we evaluate by generating predictions for the leading order case, i.e. model LO-2D, and comparing them with those of \citep{SlepianEisenstein2017} (model LO-1D). The differences between the two cases are well below the 1-$\sigma$ Gaussian uncertainty.

    \item Differences between the next-lo-leading and the leading order model predictions depend on the scale and on the redshifts, as expected. The two sets of models agree with each other on large scales. The matching scale, defined as the one in which the differences between model predictions are smaller than the expected Gaussian error, being larger at $z=0.49$ than at $z=1.05$.
    On small scales, the NLO model performs better than the LO one. To quantify the significance of the improvement, we have compared the residuals of the two model predictions with respect to the 3PCF measured in the snapshots of the  DEMNUni simulation. At $z=0.49$, the NLO mean residuals are significantly smaller than the LO ones, the statistical significance being at the 4-$\sigma$ level, where the {\it rms} scatter has been estimated from the simulations. At $z=1.05$, an improvement is also seen, though at a 2-$\sigma$ significance level.
    The NLO model improvement is even more evident on small scales, i.e. on triangle configurations in which $r_{12}\leq 40 \, h^{-1} \mathrm{Mpc}$. On those scales, the significance of the residual difference is as large as 6-$\sigma$ at $z=0.49$. 
    

    \item The 3PCF \MG{NLO-SPT and NLO-EFT} models outperform the LO one also on the BAO scale, whose importance for cosmological analyses cannot be overstressed. 
    In particular, we focused on isosceles squeezed BAO configurations in which the smallest triangle side is $r_{12} = \MG{22.5} \ h^{-1} \mathrm{Mpc} $ to probe the nonlinear regime. The quantitative comparison reveals that the reduced $\chi^2$ of the LO model is \MG{2} times larger than the \MG{NLO-SPT} one at $z=0.49, 1.05$ \MG{, and 4 and 10 times larger at $z=0.49, 1.05$, respectively.} 
    \MG{However, from a visual inspection, neither models seem to perfectly match the shape (and the position) of the BAO peak in the 3PCF, but the correlation between measurements and errors makes it difficult to quantify the significance of the mismatch.}
    
        \item In this work, we have considered two next-to-leading models. The first one,  whose performances have been described so far, is based on standard perturbation theory and dubbed NLO-SPT. The second one is based instead on the effective field theory of the large-scale structure, and it is dubbed NLO-EFT. This second model depends on four free parameters that we do not fix, but we determine by minimizing the $\chi^2$ function, and we do this separately for both redshift and for each choice of $r_{\mathrm{min}}$, the smallest size in each triangle configurations where the other two sizes  span the range  $[17.5, \MG{132.5}] h^{-1} \mathrm{Mpc}$. Moreover, in our analysis, we explored two sets of triangle configurations: the first one labelled $\eta_{\mathrm{min}} = 0$ that includes all triangles, and the second one, $\eta_{\mathrm{min}} = 2$, in which $|r_{13}-r_{12}|\geq 10 h^{-1} \mathrm{Mpc}$, EFT free parameters are estimated for these two cases separately too.
        It turns out that the best-fit EFT parameters are significantly different from zero only \MG{on the small scales range and larger scales starting approaching the BAO regime}, and the precision of their estimate improves from $\eta_{\mathrm{min}} = 2$ to $\eta_{\mathrm{min}} = 0$.

        \item To assess the relative performance of the three models, NLO-SPT, NLO-EFT, and LO-SPT, we estimate and compare their reduced $\chi^2$ difference with respect to the DEMNUni measurements as a function of $r_{\mathrm{min}}$. 
        For the case $\eta_{\mathrm{min}} = 2$, the analysis of the cumulative $\chi^2$ functions shows that all models perform similarly on scales larger than $\MG{40} h^{-1} \mathrm{Mpc}$, whereas on smaller scales both next-lo-leading order models, NLO-SPT and NLO-EFT - this latter slightly better than the former -, outperform LO-SPT.
        However, this occurs on scales in which no model provides a satisfactory good fit to the data, although the quality of the fit is assessed assuming Gaussian errors, which surely underestimates the true uncertainty on these nonlinear scales.
        A similar conclusion holds true for $\eta_{\mathrm{min}} = 0$, although in this case, the scale at which the linear order model starts underperforming with respect to the NLO ones is as large as 
        $\MG{40} h^{-1} \mathrm{Mpc}$. 

        \item Focusing on the relative performance of NLO-SPT and NLO-EFT, we find that the two models perform 
        almost identically \MG{at $\eta_{\mathrm{min}} = 2$}. \MG{For $\eta_{\mathrm{min}} = 0$, when all triangle configurations are considered, the NLO-EFT model provides a better fit to the data than NLO-SPT in the range below $r_{\mathrm{min}} = 50 h^{-1} \mathrm{Mpc}$}.

\end{itemize}

This work represents a first step towards a next-to-leading model for the 3-point correlation function of galaxies, rather than matter, to be compared with those measurements that will be performed on next-generation datasets containing tens of millions of objects. We have shown that the use of the 2D-FFTLog technique allows us to design a numerical model for the 3PCF, which, though computationally intensive, can be generated using standard computing facilities.
We have performed a number of tests to validate the procedure, to gauge the model improvement over the existing leading order one and to assess its ability to actually reproduce the 3-point correlation function of the dark matter particles in state-of-the-art N-body simulations.

We are now ready for the next step and generate a next-to-leading order  3-point correlation function of the galaxies by
including a non-linear, non-local bias relation between galaxies and matter. This will be done following the same strategy proposed in this work, i.e. the use of the 2D-FFTLog technique to transform the various terms that define the next-to-leading order model for the galaxy bispectrum \cite{Eggemeier2019, Eggemeier2021} to configuration space and combine them to obtain the prediction for the galaxy 3PCF. A similar suite of validation tests will be performed using catalogues of dark matter halos extracted from N-body experiments.

A second, equally important effort needs to be made before comparing the 3PCF model to the actual data to constrain cosmological parameters: the assessment of the 3PCF errors and their covariance. Clearly, a pure brute-force approach based on the numerical evaluation of the covariance matrix is unfeasible since it would require measuring the 3-point correlation function for a large number of objects and triangle configurations in a large suite of realistic mock catalogues.
A more realistic strategy based on a hybrid approach that combines theoretical errors with numerically-evaluated ones will have to be defined together with a sensible choice of the triangle configurations that will allow for maximising the scientific return of the analysis.

Finally, because modelling the 3PCF is computationally expensive, standard MCMC techniques will probably be too slow to efficiently sample the posterior probability distribution in the high dimensional space of the free parameters once nonlinear galaxy bias recipes will be included in the model. A faster strategy, e.g. based on the efficient implementation of an emulator technique will have to be adopted.

\acknowledgments
\MG{We thank the anonymous referee for the suggestions that helped improve the paper}. We also thank Michele Moresco, Kevin Pardede and Anna Pugno for useful discussions.
This work is supported by MIUR/PRIN 2017 ``From Darklight to Dark Matter: understanding the galaxy-matter connection to measure the Universe'', the INFN project ``InDark'' and 
by ASI/INAF agreement n. 2018-23-HH.0
``Scientific activity for Euclid mission, Phase D''. AE is supported at the AIfA by an Argelander Fellowship.
The DEMNUni simulations were carried out in the framework of ``The Dark Energy and Massive-Neutrino Universe'' project, using the Tier-0 IBM BG/Q Fermi machine and the Tier-0 Intel OmniPath Cluster Marconi-A1 of the Centro Interuniversitario del Nord-Est per il Calcolo Elettronico (CINECA). We acknowledge a generous CPU and storage allocation by the Italian Super-Computing Resource Allocation (ISCRA) as well as from the coordination of the ``Accordo Quadro MoU per lo svolgimento di attività congiunta di ricerca Nuove frontiere in Astrofisica: HPC e Data Exploration di nuova generazione'', together with storage from INFN-CNAF and INAF-IA2.

\setlength{\bibsep}{2pt plus 0.5ex}
\bibliographystyle{JHEP}
\nocite{*}
\bibliography{bibliography}

\appendix
\newcommand{\de}{\mathrm{d}}
\renewcommand{\vec}{\mathbf}
\newcommand{\med}[1]{\left \langle #1 \right \rangle}
\newcommand{\be}{\begin{equation}}
\newcommand{\ee}{\end{equation}}
\newcommand{\bea}{\begin{eqnarray}}
\newcommand{\eea}{\end{eqnarray}}
\newcommand{\bdm}{\begin{displaymath}}
\newcommand{\edm}{\end{displaymath}}
\newcommand{\nn}{\nonumber}
\newcommand{\xv}{\mathbf{x}}
\newcommand{\yv}{\mathbf{y}}
\newcommand{\kv}{\mathbf{k}}
\newcommand{\nv}{\mathbf{n}}
\newcommand{\mv}{\mathbf{m}}
\newcommand{\rv}{\mathbf{r}}
\newcommand{\qv}{\mathbf{q}}
\newcommand{\pv}{\mathbf{p}}
\newcommand{\del}{\delta}
\newcommand\td{\widetilde{\delta}}
\newcommand{\Le}{{\mathcal L}}
\newcommand{\intq}{{\int_{\qv}}}
\newcommand{\kmin}{k_{\rm min}}
\newcommand{\kmax}{k_{\rm max}}

\section{Perturbation Theory in Fourier space}

The expression of the one-loop matter bispectrum in SPT is the following
\be
B_m(k_1,k_2,k_3)=B_m^{\rm tree}+B_m^{\rm 1-loop}+B_m^{\rm ctr},
\ee
where 1-loop corrections are given by
\be
    B_m^{\rm 1-loop} = B_{222}^{\rm 1-loop}+B_{321,I}^{\rm 1-loop}+B_{321,II}^{\rm 1-loop}+B_{411}^{\rm 1-loop}\,,
\ee
where
\bea
B_{222}^{\rm 1-loop}
& = &
8\,\intq\, F_2(-\qv,\kv_3+\qv) 
 F_2(\kv_3+\qv,\kv_2-\qv)\,F_2(\kv_2-\qv,\qv)\,P_L(q)\, P_L(|\kv_2-\qv|)\,P_L(|\kv_3+\qv|) 
\label{eq:b222}
\\
B_{321,I}^{\rm 1-loop}
& =&  6\,P_L(k_3)\intq \,F_3(-\qv,-\kv_2+\qv,-\kv_3)\,F_2(\kv_2-\qv,\qv)\,P_L(|\kv_2-\qv|) 
\, P_L(q)
+ \text{5 perm.}
\label{eq:b321i}
\\
B_{321, II}^{\rm 1-loop}
& = & 
6\, P_L(k_2)\, P_L(k_3)\,F_2(\kv_2,\kv_3) \intq\,F_3(\kv_3,\qv,-\qv)\,P_L(q) + \text{5 perm.}
\label{eq:b321ii}
\\
B_{411}^{\rm 1-loop}
& =  & 
12\,P_L(k_2)\,P_L(k_3) \intq\, F_4(\qv,-\qv,-\kv_2,-\kv_3)\,P_L(q) + \text{2 perm.} \, ,
\label{eq:b411}
\eea
where 

\begin{align}
    F_{n}(&\textbf{q}_1, ..., \textbf{q}_n ) = \\ & = \sum_{m = 1}^{n-1} \frac{G_m (\textbf{q}_1, ..., \textbf{q}_n )}{(2n+3)(n-1)} [3\alpha(\textbf{k}_1, \textbf{k}_2) F_{n-m}(\textbf{q}_{m+1}, ..., \textbf{q}_n ) + 2n\beta(\textbf{k}_1, \textbf{k}_2)G_{n-m}(\textbf{q}_{m+1}, ..., \textbf{q}_n )]
\end{align}
where $\textbf{k}_1 = \textbf{q}_1 + .. + \textbf{q}_m$, $\textbf{k}_2 = \textbf{q}_1 + .. + \textbf{q}_m$, and $F_1$ = $G_1$ = 1. For $n = 2 $ we have: 

\begin{align}
    F_2(\textbf{q}_1, \textbf{q}_1) = \frac{5}{7} + \frac{\textbf{q}_1 \cdot \textbf{q}_1}{q_1 q_2}(\frac{q_1}{q_2} + \frac{q_2}{q_1}) + \frac{2}{7} \frac{(\textbf{q}_1 \cdot \textbf{q}_2)^2}{q_1^2 q_2^2}, \\
    G_2(\textbf{q}_1, \textbf{q}_1) = \frac{3}{7} + \frac{\textbf{q}_1 \cdot \textbf{q}_1}{q_1 q_2}(\frac{q_1}{q_2} + \frac{q_2}{q_1}) + \frac{4}{7} \frac{(\textbf{q}_1 \cdot \textbf{q}_2)^2}{q_1^2 q_2^2}. 
\end{align}
To compute EFT bispectrum contribution we use the $\Tilde{F}_2^{(\mathrm{s})}(\textbf{k}_1, \textbf{k}_2)$ kernel defined as \citep{Angulo:2014tfa, Baldauf2015}


\begin{equation}
    \begin{split}
    \Tilde{F}_2^{(\mathrm{s})}(\textbf{k}_1, \textbf{k}_2) = & \frac{1}{(1 + \zeta)(7 + 2\zeta)} [(5 + \frac{113\zeta}{14} + \frac{17 \zeta^2}{7})(k_1^2 + k_2^2) \\
    & + (7 + \frac{148\zeta}{7} + \frac{48\zeta^2}{7})(\textbf{k}_1 \cdot  \textbf{k}_2)  \\
    & + (2 + \frac{59\zeta}{7} + \frac{18\zeta^2}{7}) (\frac{1}{k_1^2} + \frac{1}{k_2^2}) (\textbf{k}_1 \cdot  \textbf{k}_2)^2 \\
    & + (\frac{7}{2} + \frac{9\zeta}{2} + \zeta^2)(\frac{k_1^2}{k_2^2} +\frac{k_2^2}{k_1^2})(\textbf{k}_1 \cdot  \textbf{k}_2) \\
    & + (\frac{20\zeta}{7} + \frac{8\zeta^2}{7}) \frac{(\textbf{k}_1 \cdot  \textbf{k}_2)^3}{k_1^2k_2^2} ].
    \end{split}
\end{equation}
Following \cite{Angulo:2014tfa}, we fix $\zeta = 3.1$, but we note that in \citep{Baldauf2015} there is no evidence of different predictions from adopting $\zeta = 2$. 
\label{AppendixA}

\section{2D-FFTLog technique}
\label{AppendixC}
\MG{The implementation of the 2D-FFTLog procedure to estimate the 3PCF relies on } the dimensionless bispectrum multipoles:


\begin{equation}
    \label{eq:dimless_bisp}
    \Delta_\ell(\vec{k}_{12}, \vec{k}_{13}) = \frac{k_{12}^3 k_{13}^3}{(2\pi^2)^2} B_\ell(\vec{k}_{12}, \vec{k}_{13}). 
\end{equation}
\MG{that are used to estimate the 3PCF multipoles $\zeta_\ell$ from Eq. \ref{eq:zeta_from_bkell}}.
Using for $\Delta_\ell(\vec{k}_{12}, \vec{k}_{13}) $ the power law expansion Eq.  \ref{eq:dimless_bisp} one obtains:
\begin{equation}
    \begin{split}
    \label{eq:FFTLog_explanation}
    \zeta_\ell (r_{12}, r_{13})  &=  (-1)^{\ell} \sum_{m, n = -N/2}^{N/2} c_{\ell, mn} \frac{1}{k_{1, 0}^{i\nu_m}} \frac{1}{k_{2, 0}^{i\nu_n}}  \\ 
    &\int \frac{dk}{k_1} k_{1}^{\nu_1 + i\nu_m} j_\ell(k_1r_1) \int \frac{dk}{k_2} k_{2}^{\nu_2 + i\nu_m} j_\ell(k_2r_2) = \\
     = &(-1)^{\ell} \frac{\pi}{16 r_{12}^{\nu_1} r_{13}^{\nu_2}} \sum_{m, n = -N/2}^{N/2} c_{\ell, mn} k_0^{-i(\eta_m + \eta_n)} r_{12}^{-i\eta_m} r_{13}^{-i\eta_m} \\ 
     & \times g_\ell(\nu_1 + i\eta_m) g_\ell(\nu_2 + i\eta_n),
    \end{split}
\end{equation}
where $g_\ell(\omega) = 2^\omega \frac{\Gamma(\frac{\ell + \omega}{2})}{\Gamma(\frac{3 + \ell - \omega}{2})}$. The range of validity of $\nu_1$ and $\nu_2$ are $-\ell < \nu_1, \nu_2 < 2$. 
\MG{The 2D-FFTLog method performs the 2D FFT transformation twice, and thus, its computational cost scales as $\mathcal{O}(N^2 \mathrm{log} N)$
Inserting the expression \eqref{eq:zeta_binave} into \eqref{eq:FFTLog_explanation} and re-naming the denominator as $A$ it is possible to obtain the bin-averaged expression for the 3PCF multipoles 
that is evaluated using the 
for the 2D-FFTLog algorithm as follows}
\begin{equation}
\begin{split}
\label{eq:zeta_binave_FFT}
\zeta_\ell (\bar{r}_{i}, \bar{r}_{j}) = \frac{\pi \bar{r}_{i, \mathrm{min}}^{2-\nu_1} \bar{r}_{j, \mathrm{min}}^{2-\nu_2} }{16 A N^2} \sum_{m, n = -N/2}^{N/2} c_{\ell, mn} k_0^{-i(\eta_m + \eta_n)} r_{12}^{-i\eta_m} & r_{13}^{-i\eta_m} g_\ell(\nu_1 + i\eta_m) g_\ell(\nu_2 + i\eta_n)\times \\ & \times s(D -\nu_1 -i\eta_m, \lambda) s(D -\nu_2 -i\eta_n, n)
\end{split}
\end{equation}
where
\begin{equation}
    s(D, n) = \frac{n -1}{D}, 
\end{equation}
where $\frac{\bar{r}_{i, \mathrm{max}}}{\bar{r}_{i, \mathrm{min}}} = n $ is the linear bin width, and
\begin{equation}
    A = \int_{\bar{r}_{i, \mathrm{min}}}^{\bar{r}_{i, \mathrm{max}}} dr_1 r_1^2 \int_{\bar{r}_{i, \mathrm{min}}}^{\bar{r}_{i, \mathrm{max}}} dr_2 r_2^2 = \bar{r}_{i, \mathrm{min}}^2  \bar{r}_{j, \mathrm{min}}^2 [s(2, n)]^2. 
\end{equation}

\section{2D-FFTLog calibration}
\label{AppendixB}

To correctly use FFTlog machinery, we studied the algorithm as a function of input parameters. We considered as input parameters $k_{\mathrm{min}}, k_{\mathrm{max}}$, the minimum and maximum values over we perform Eq \ref{eq:zeta_from_bkell}, a damping factor $k_0$, i.e.

\begin{equation}
    P_{k_0}(k) = P_L(k) e^{-k^2/k_0^2}, 
\end{equation}
we used to study convergence properties and ringing sensibility of 1D and 2D FFTlog algorithms, $N_k$ and $N_\mu$ numbers of point in k-direction (1D and 2D) in $\theta$-direction (only 2D). 

\begin{figure}[ht!]
    \includegraphics[width=1.\textwidth]{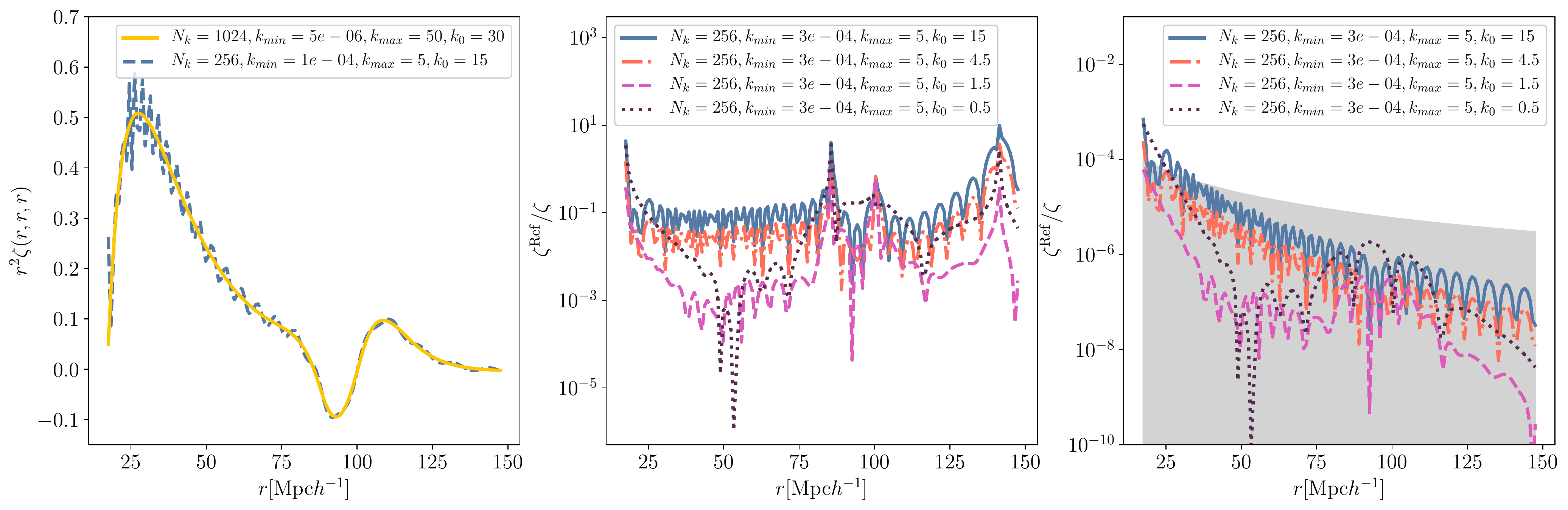}
    \caption{Different computation of $\zeta$ for equilateral computation with different grid parameters using 1D-FFTLog. Left: plot of different evaluations. Middle: ratio of previous computations. Right: residuals of previous evaluations.} 
    \label{fig:appendix_1D}
\end{figure}

In Fig. \ref{fig:appendix_1D}, we tested, at $z=0.5$, different usage of parameters reproducing the LO-1D model at the centre of the bin and considering an extended k-range from $k_{\mathrm{min}} = 5 \times 10^{-6}$ and $k_{\mathrm{max}} = 5 \times 10^1$ as a reference model. An incorrect using of damping $k_0$ over a shorter grid implies a source of ringing effect. Using $k_0 = 1.5  h \mathrm{Mpc}^{-1}$ is the best solution to mimic a usage of a larger grid and minimise ringing effects. 

\begin{figure}[H]
    \includegraphics[width=1.\textwidth]{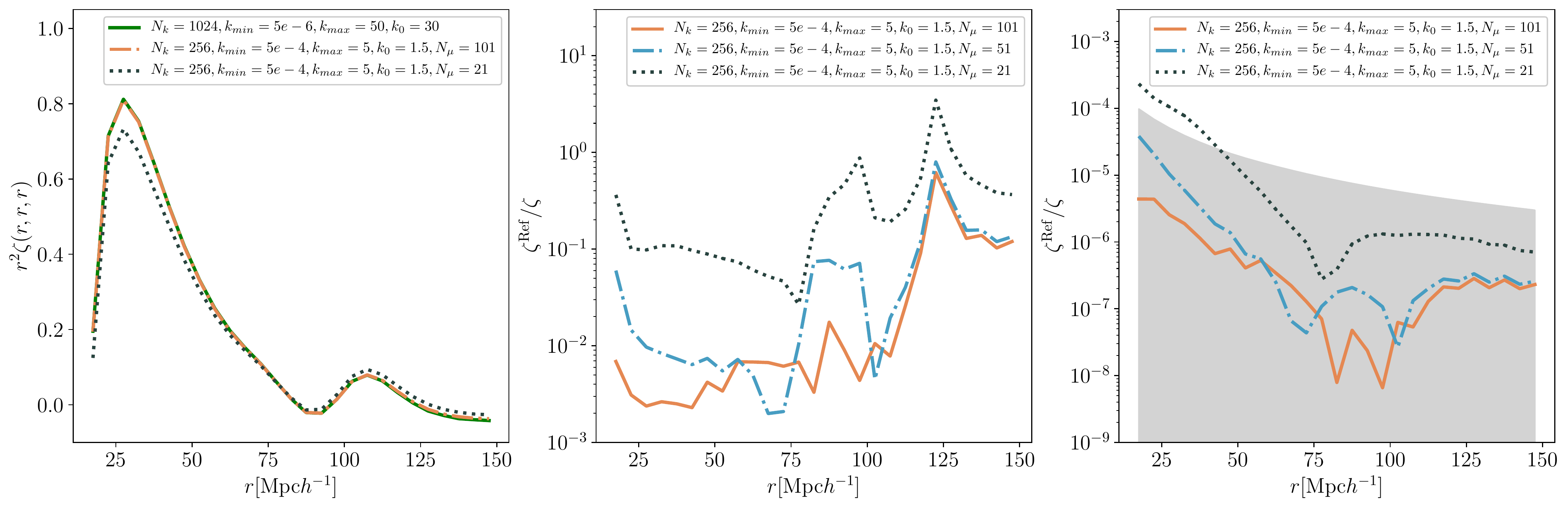}
    \caption{Different computation of $\zeta$ for equilateral computation with different grid parameters using 2D-FFTLog. Left: plot of different evaluations. Middle: residuals of previous computations. Right: difference between previous evaluations compared to theoretical error assuming theoretical covariance.} 
    \label{fig:appendix_2D}
\end{figure}

In Fig. \ref{fig:appendix_2D} we did the same test but using the 2D version of the FFTLog algorithm, considering the LO-2D model, including several sample points in $\theta$-direction but, differently from the previous case, considering the bin-averaged model. In this case, we fix the damping factor according to the previous case $k_0 = 1.5 h \mathrm{Mpc}^{-1}$ where
\begin{equation}
    B_{k_0}(k_1, k_2, k_3) = B(k_1, k_2, k_3) e^{-(k_1^2 + k_2^2 + k_3^2)/k_0^2}, 
\end{equation}
We found $N_{\theta}$ needs to be at least equal to 51 to minimise the difference with the reference model (1D-FFTlog evaluation). We use $N_{\theta} = 51$, $N_{k} = 256$, $k_{\mathrm{min}} = 5 \times 10^{-6}$ and $k_{\mathrm{max}} = 5 \times 10^1$ in our computations.

\section{The Slepian-Eisenstein matter 3PCF}
\label{AppendixD}

Considering the leading order, or the tree-level bispectrum in Eq. \eqref{eq:B_tree} (see \citep{Bernardeau2002}), it is possible to obtain the multipoles of the 3PCF after some manipulations on \eqref{eq:zeta_from_bkell}, performing analytical integrations over angular variables as presented in \citep{SlepianEisenstein2017}:

\begin{equation}
\begin{split}
\zeta_\ell(r_{12}, r_{13})  = \frac{2 \ell+1}{2} & \int_{-1}^{1} \mathrm{~d} \mu_{12} [ \zeta_{\mathrm{pc}}(r_{12}, r_{13}, \mu_{23}) +
\\ & + \zeta_{\mathrm{pc}}(r_{13}, r_{23}, \mu_{23})+\zeta_{\mathrm{pc}} (r_{23}, r_{12}, \mu_{13}) ] \mathcal{P}_\ell(\mu_{23}),
\end{split}
\end{equation}
where $\mu_{23} = \frac{\mathbf{r_{12} \cdot r_{23}}}{r_{12}r_{23}}$ and similarly for $\mu_{13}$, $\mu_{12}$. The $\zeta_{\mathrm{pc}}$ terms are the pre-cyclic contributions to matter 3PCF, computed as:
\begin{equation}
\label{eq:zeta_semianalytic}
\zeta_{\mathrm{pc}}(r_{12}, r_{13} ; \hat{r}_{12} \cdot \hat{r}_{13})=\sum_{\ell=0}^{2} \zeta_{\mathrm{pc, \ell}}(r_{12}, r_{13}) P_\ell(\hat{r}_{12} \cdot \hat{r}_{13})
\end{equation}
where the Legendre coefficients are:
\begin{align}
    \zeta_{\mathrm{pc}, 0}(r_1, r_2) & = \frac{34}{21} \xi_1 \xi_2, \\
    \zeta_{\mathrm{pc}, 1}(r_1, r_2) & = -[\xi^{[1+]}_1 \xi^{[1-]}_2  + \xi^{[1-]}_1 \xi^{[1+]}_2 ], \\
    \zeta_{\mathrm{pc}, 2}(r_1, r_2) & = \frac{8}{21} \xi^{[2]}_1 \xi^{[2]}_2.
\end{align}
The $\xi$ terms are 1D integrals of
the input power spectrum:
\begin{align}
    \label{eq:xi_SE_1}
    \xi_i^{[n]}(r) & = \int_0^{\infty} \frac{\mathrm{d}k}{2\pi^2} k^2 j_n (k r_i) P(k), \\
    \label{eq:xi_SE_2}
    \xi_i^{[n\pm]}(r) & = \int_0^{\infty} \frac{\mathrm{d}k}{2\pi^2} k^2 k^{\pm 1} j_n (k r_i) P(k)
\end{align}
where $j_l$ are the spherical Bessel functions.

\end{document}